\documentclass[a4paper]{article}
\usepackage{a4}
\usepackage[compact]{titlesec}

\usepackage{longtable}
\usepackage{amssymb}

\def\xcite#1{\cite{#1}}

\begin{document}

\begin{center}
{\bf\Large 
The Encyclopedic Reference of Critical Points for $SO(8)$-Gauged
  $\mathcal{N}=8$ Supergravity\\[2ex]
Part 1: Cosmological Constants in the Range $-\Lambda/g^2\in[6;14.7)$}

\bigbreak

{\bf Thomas Fischbacher\\}
\smallbreak
{\em University of Southampton\\
  School of Engineering Sciences\\
  Highfield Campus\\
  University Road, SO17 1BJ Southampton, United Kingdom\\}
{\small {\tt t.fischbacher@soton.ac.uk}}

\end{center}

\begin{abstract}
  \noindent This article is part of a collection that strives to
  collect and provide in an unified form data about all the critical
  points on the scalar manifold of $SO(8)$-gauged $\mathcal{N}=8$
  supergravity in four dimensions known so far. The vast majority of
  these were obtained using the enhanced sensitivity backpropagation
  method introduced by the author in 2008.

  This part of the collection describes 41 critical points, 7 of which
  have been known for more than two decades, 8 of which were
  discovered recently, and 26 are novel. The residual gauge symmetries of
  these 41 critical points (likely) are SO(8) with N=8 SUSY (1x),
  SO(7) (2x), SU(4) (1x), G2 with N=1 SUSY (1x), SU(3)xU(1) with N=2
  SUSY (1x), SO(3)xSO(3) (2x), SO(3)xU(1)xU(1) (1x), SO(3)xU(1) (3x),
  SO(3) (3x), U(1)xU(1) with N=1 SUSY (1x), U(1)xU(1) without SUSY (4x), U(1)
  (11x), and None (10x). Analytic conjectures (not yet proven but
  overwhelmingly likely correct) are given for the locations and
  cosmological constants of some critical points.

\end{abstract}

\setcounter{section}{-1}
\section{About this document}

\noindent Thanks to novel powerful methods introduced by the author
in~\xcite{Fischbacher:2008zu,Fischbacher:2009cj}, the symmetry breaking
structure of field theories with large numbers of scalars can now be
investigated to unprecedented depth. A systematic application of these
methods to $SO(8)$-gauged $\mathcal{N}=8$ supergravity~\xcite{de Wit:1981eq,de Wit:1982ig}
greatly expanded our knowledge of locations and properties of critical
points, increasing the number of known solutions of the stationarity
condition from seven to well over one hundred.

While these new methods manage to produce a far more detailed picture
than earlier approaches, they neither are exhaustive, nor do they give
analytical results straightaway. (It is indeed possible to obtain
analytic expressions in a fully automated way, as has been
demonstrated for some novel critical points, but this often requires
considerable computing time.) Hence, it is natural to expect that many
more details will become known about this previously mostly uncharted
territory, the scalar sector of gauged maximal four-dimensional
supergravity. These critical points are interesting for a number of
reasons, not least because we think to have some useful degree of
control over the mathematical techniques needed to work with
supergravity backgrounds.

Traditionally, the discovery and detailed discussion of one or a few
new solutions of the stationarity conditions would have warranted a
dedicated research article, cf.
e.g.~\xcite{Pernici:1984zw,Warner:1983du,Khavaev:1998fb,Hull:1984ea}. If
the novel methods now available would have led to a mere doubling of
the number of known critical points, it may have been appropriate to
discuss them individually in detail in one article each. Now that the
supergravity scalar potential turned out to have a much more
complicated structure than what one may have anticipated, the question
of \emph{organizing} and presenting the data in a useful way becomes
quite relevant. Pondering over the detailed list, one can make a few
remarkable observations; often, there are pairs of critical points
with quite similar properties, sometimes perhaps related to one
another by triality, sometimes not (or not obviously) so. Clearly,
such things would be very difficult to spot were all this information
scattered over many individual articles, each perhaps using different
conventions for the parametrization of some particular scalar
sub-manifold under study. Therefore, a compact and unified overview
presentation of all results makes sense.

\subsection{Purpose}

The purpose of this series, initially consisting of four parts, is
to:

\begin{itemize}

\item Collect information about critical points of maximal
  $SO(8)$-gauged supergravity in four dimensions that have become known.

\item Provide (at least on the arXiv) an up-to-date reference listing
  solutions, properties, and research articles where they are used or
  investigated further in a systematic way.

\end{itemize}

It is planned to provide updates to the master list on arXiv whenever
new solutions become known, or relevant new properties of specific
solutions are discovered, accompanied by separate overviews over
updates after relevant major new discoveries.

In order to make this update process manageable and accessible:

\begin{itemize}

\item The complete list has been split into parts of manageable size,
  each containing about~40 solutions in the version~1 release. Each
  part can receive independent updates and grow independently from one
  another, using independent version numbers.

\item The discussion of properties of solutions carefully is worded in
  such a way that global statements which may be invalidated by
  further discoveries (such as ``this is the only known solution with
  property X'') are avoided. This ensures that each part can evolve
  without the need for ever deleting statements, or for modifying
  statements that involve different parts of the collection.

\end{itemize}

This means, of course, that global statements such as ``there is only
one known non-supersymmetric solution that satisfies the
Breitenlohner-Freedman stability condition'' (true at the time of the
release of the collection of `version~1' parts) cannot be made. This
should be kept in mind when using the list.

A ``big picture'' overview of the solutions contained in the initial
release of the collection will be given in a separate
article~\xcite{Fischbacher2011inprep}.

Two questions naturally arise here: One may wonder whether the number
of solutions actually might be so large that such a classification
effort might be futile. Also, the question how the claimed properties
of such a large number of solutions can be validated is quite
relevant.

Concerning the expected total number of solutions, including yet
undiscovered ones, the raw data produced by the deep search that
resulted in the greatly enlarged list contained numerous duplicates;
while the methods used so far seem to systematically miss certain
solutions, it also now appears somewhat unlikely for the total number
of critical points to be larger than 1000 or so.

The validation problem is addressed by the
article~\xcite{Fischbacher:2010ki}. The arXiv source code of that
article, available at \texttt{http://arxiv.org/e-print/0912.1636}, contains a
direct Python transcription of the definitions presented in detail
in~\xcite{Fischbacher:2009cj} that can be used to numerically check all
claims about critical points. Raw numerical data files for each of the
published solutions can be obtained by downloading the source file of
the corresponding part of the collection from arXiv. The numerical
accuracy provided by these data files considerably goes beyond the
number of digits actually listed in the text.

\subsection{Structure of the tables}

Each table lists, in its headline, an boldface identifier that
uniquely names the critical point. As it is expected that new
solutions will be discovered that fall between known ones, these
identifiers cannot be consecutive numbers. As the cosmological
constant turns out to be a useful proxy to discern solutions, these
names are based on the value of the potential at the critical
point. In most cases, the name of a stationary point consists of the
letter `S' followed by a seven-digit code which is the {\em truncated}
(rather than rounded) cosmological constant multiplied by $-10^5$. The
headline furthermore lists the value of the potential to somewhat
greater accuracy, and the base-10 logarithm of the Frobenius norm
square of the `Q-tensor' (which gives the violation of the
stationarity condition) for the numerical solution. This roughly
matches the number of valid digits in the cosmological constant; as
the potential is quadratic around a stationary point, the number of
valid digits in the $\phi$ coordinates is roughly half the number of
valid digits in the potential. Furthermore, the headline lists
information about the dimension of the residual unbroken gauge group
(if any -- many solutions do not have any residual symmetry), and
possible the amount of supersymmetry.

The headline is followed by a block listing the non-zero coordinates
of the 70-dimensional vector $\phi$. These have been converted from
the $35_s+35_c$ form used in~\xcite{Fischbacher:2009cj} to the language of
self-dual/anti-self-dual 4-forms. A term such as $-0.201180_{1238s}$
e.g. means that the corresponding contribution to the symmetric
traceless matrices over the spinors (index tag `s') is given by
$-0.201180\cdot\gamma^{1238}_{\alpha\beta}$. Users of these tables may
find both the validation code given in~\xcite{Fischbacher:2010ki}
as well as data describing previously known critical points useful to
match their conventions against the ones used here. As discussed
in~\xcite{Fischbacher:2008zu,Fischbacher:2009cj}, the minimization of $|Q|^2$
(the violation of the stationarity condition) via sensitivity
backpropagation produces a solution in random position on a gauge
group orbit. A second optimization step is used to find a gauge group
rotation that sets as many coordinates to zero as possible, giving
both a nice presentation with substantially fewer than~70 nonzero
coordinates, as well as numerically suggesting simplifying relations
between coefficients. While this second optimization step usually is
much faster than the first, there are some odd cases where it has a
tendency to `get stuck', resulting in a presentation with many more
nonzero entries than necessary. It is quite possible that some (few)
of the solutions listed in the tables suffer from this problem.

Finally, each table contains three blocks listing the (ordered)
masses-squared of gravitini $(m^2/m_0^2)[\psi]$, spin-$1/2$ fermions,
and the masses of scalars, with multiplicities. All these masses are
given in AdS units. Each unbroken supersymmetry corresponds to a
gravitino mode with mass-squared $+1$, and the
Breitenlohner-Freedman~\xcite{Breitenlohner:1982jf,Breitenlohner:1982bm}
stability criterion for scalar masses is
$(m_/m_0)[\phi]\ge-(d-1)^2/4=-2.25$.

While these tables give a first numerical overview over properties of
solutions, and all numbers shown are expected to be correct to as many
digits as are actually listed, it must again be pointed out that the
raw source files (obtainable from arXiv via the `other formats'
article download format link) contain more accurate data than what is
shown in print. These hence may be useful for some investigations.

\section{Introduction}

Supersymmetry~\xcite{Wess:1974tw} is a highly nontrivial extension of
space-time symmetries with far reaching consequences. As supersymmetry
transformations mix bosonic and fermionic degrees of freedom,
supersymmetric field theories are constrained both in their particle
spectra as well as in the form of their interaction terms. These
constraints become stronger as one goes from minimally supersymmetric
field theories to models with extended supersymmetry. One finds that
it is possible to not only construct supersymmetric field theories in
flat space, but also to supersymmetrize general relativitiy, which
leads to supergravity~\xcite{Freedman:1976xh,Deser:1976eh}. As one
can obtain a spacetime translation from two supersymmetry
transformations, a supersymmetric theory including general relativity
must necessarily feature supersymmetry as a {\em local} symmetry,
which then gives rise to diffeomorphisms rather than constant
translations (which in general do not exist in curved spacetime). Both
the attempt to promote supersymmetry to a local (gauge) symmetry as
well as the attempt to combine supersymmetry with general relativity
hence lead to the same goal, namely supergravity. Therefore, should
supersymmetry be proven experimentally to play a role in particle
physics, then accepting general relativity as the appropriate
classical description of gravity inevitably forces us to accept the
idea of supergravity. Just as there are non-gravitational field
theories with extended (i.e. more than minimal) supersymmetry, there
also are supergravities with extended supersymmetry.

The simplest supergravity features a spin-2 graviton as well as a
spin-3/2 gravitino. Each additional supersymmetry enlarges the
supermultiplet helicity range by 1/2, up to a maximum of eight times
the minimal amount of supersymmetry. In this maximally supersymmetric
supergravity, the helicity +2 graviton state is unified with all other
particle states down to the helicity -2 graviton state into a single
supermultiplet whose structure is completely fixed by
supersymmetry. As gravitational theories with more than the minimal
number of graviton states do not seem to make sense, and it seems
impossible to construct theories of interacting massless particles
with helicity larger than 2, the maximal amount of supersymmetry in a
3+1-dimensional field theory with gravity is constrained to be
$\mathcal{N}=8$ times the minimal amount of supersymmetry, with 32
supersymmetries in total (eight real four-spinors).

Considering the particle content of this `$\mathcal{N}=8$
supergravity', which is completely fixed by supersymmetry, there are
one graviton, eight gravitini $\psi$, 28~vectors~$A$, 56~spin-1/2
fermions~$\chi$, and finally a total of
70~scalars~$\phi$. Historically, the construction of the Lagrangian
for $\mathcal{N}=8$ supergravity did not succeed by using the
iterative procedure (the Noether method) employed for models with less
supersymmetry, due to the complicated interactions between the
scalars. Ultimately, the construction succeeded by a detour employing
a Kaluza-Klein reduction~\xcite{Duff:1986hr}: first, a supergravity was
constructed in the maximum possible spacetime dimension that permits
fermion-boson matching (and has one time direction). This gave rise to
eleven-dimensional supergravity~\xcite{Cremmer:1978km}, which, when
compactified on a 7-torus $T^7$ to four spacetime dimensions, gives
the desired $\mathcal{N}=8$ supergravity~\xcite{Cremmer:1979up}. A
surprising discovery was that generalized electric-magnetic duality
plays a very prominent role for this theory and involves a non-compact
real form of one of the exceptional simple Lie groups in the
Cartan-Dynkin classification, the group $E_{7(7)}$. Specifically, the
scalars of the theory parametrize the 70-dimensional riemannian
symmetric space $E_{7(7)}/SU(8)$. While this, like 11-dimensional
supergravity, was originally regarded mostly as a mathematical
curiosity, the modern perspective considers 11-dimensional
supergravity to play a central role as the low energy limit of an at
the time of this writing only partially understood `mother' theory
(`$\mathcal{M}$-Theory') which, in different limits, gives rise to the
five 10-dimensional superstring theories (and supergravities in
various dimensions)~\xcite{Witten:1995ex}. The emergence of the
$E_{7(7)}$ symmetry group is now considered to be a consequence of
duality symmetries, already present in higher dimensions, of
$\mathcal{M}$-Theory. It is likely that the most promising strategy to
obtain a deeper understanding of the structure and properties of
$\mathcal{M}$-Theory is to focus on these `U-duality'
symmetries~\xcite{Obers:1998fb,West:2004kb}.

Closer investigation of $\mathcal{N}=8$ supergravity in
four-dimensional spacetime has shown that maximal supersymmetry does
not fix the construction completely; it is possible to promote the
28~vector bosons to gauge bosons of a non-abelian symmetry, in
particular $SO(8)$~\xcite{de Wit:1982ig}. This corresponds to a 
compactification of 11-dimensional supergravity on the
7-sphere~$S^7$. 

Surprisingly, it was later found that it is just as well possible to
alternatively introduce certain \emph{non-compact} nonabelian gauge
groups. In particular, it is e.g. possible to use non-compact real
forms $SO(p,8-p)$ instead of $SO(8)$. In supergravity, this works (and
does not introduce ghosts) because the kinetic term for the vectors is
not of the form $\sim\delta_{AB} F^A F^B$ ($A, B$ being gauge group
adjoint representation indices), but of the form $\sim S_{AB}(\phi)
F^A F^B$ with some positive definite symmetric inner product $S$ that
depends on the scalars~\xcite{Hull:1983yf,Hull:1984qz}.

In the 80s, the construction of a fully unified theory of matter and
force particles including gravity that was pretty much uniquely
determined by its symmetry properties gave rise to considerable
enthusiasm. This made some authors (most notably Stephen Hawking in
his inaugural lecture as the Lucasian Professor of Mathematics at
Cambridge~\xcite{Hawking}) proclaim that we would, in the
not-too-distant future, have a Unified Theory of
Everything. Considering attempts to link $\mathcal{N}=8$ supergravity
with experimentally observed particle physics immediately face a
number of problems; perhaps most importantly, it is not possible to
accommodate the standard model gauge group $SU(3)\times SU(2)\times
U(1)$ in $SO(8)$. For this and other reasons, present interest in
$\mathcal{N}=8$ supergravity is more due to its remarkable
mathematical properties, as well as its relation to three-dimensional
CFTs via the AdS/CFT correspondence~\xcite{Maldacena:1997re} than due
to it being a viable theory of quantum gravity containing the Standard
Model. While initial hopes that supergravity may directly give
renormalizable quantum theory of gravity via boson-fermion loop
cancellations turned out to be too optimistic, the renormalizability
question is not settled yet for $\mathcal{N}=8$
supergravity~\xcite{Bern:2006kd,Kallosh:2011dp,Beisert:2010jx}. It is
already clear that $E_7$ symmetry plays a crucial role for the
question whether $\mathcal{N}=8$ supergravity may make sense in the UV
limit.

Whenever some symmetry that rotates supersymmetry generators into one
another is gauged, extra terms have to be added to the Lagrangian. In
particular, a potential for the scalar arises in second order of the
gauge coupling constant. This scalar potential is a quadratic function
of the mass matrices (and couplings) of the fermionic fields. In the
simplest case, this potential is just a cosmological
constant~\xcite{Freedman:1976aw}, but for theories with more
complicated scalar sectors, it can be fairly involved. Both the
possibility to gauge extended supergravities as well as the scalar
potential arising from this procedure are general features of extended
supergravities in various dimensions. These potentials have a number
of properties that at first seemed rather awkward; they give rise to a
(typically negative) Planck-scale cosmological constant, are unbounded
from below, and stationary points normally are saddle points (or, in
some cases, maxima), rather than minima. However, it has been shown
that even saddle points can lead to (at least perturbatively) stable
vacua, as one has to carefully analyze the interplay between kinetic
and potential energy in a negatively curved background
spacetime~\xcite{Breitenlohner:1982jf,Breitenlohner:1982bm}.

It is easy to see why conventional approaches to analyze the
potentials of gauged supergravities with a large number of scalars
cannot succeed: one would first have to construct an analytic
parametrization of a high-dimensional coset manifold such as
$E_{7(7)}/SU(8)$, then use this parametrization to express the
potential in terms of (hyperbolic) sines and cosines of the
coordinates, and finally solve a complicated set of coupled polynomial
equations. Even the first step -- parametrizing the coset manifold --
would generate, in the $E_7$ case, a complex $56\times 56$ matrix with
entries that are polynomial in (hyperbolic) sines/cosines of
70~variables. As a substantial fraction of all conceivable
combinations of factors does indeed arise, one would guesstimate the
order of magnitude of the number of terms in each matrix entry to be
around~$2^{70}\sim10^{21}$ -- certainly well beyond reach. It is
nevertheless possible to modify this approach in such a way that it
becomes analytically feasible, but the price to be paid is that the
analysis is constrained to solutions with a large amount of residual
gauge symmetry. Basically, the idea is that if $H$ is a subgroup of
the gauge group $G$ and $M_H$ is a $H$-invariant submanifold of the
scalar manifold, then any critical point $P_H$ on $M_H$ must also be a
critical point on the full scalar manifold, as the first-order term in
the potential expanded around $P_H$ is $G$-invariant and there is no
way to form a $G$-singlet from a `1-element tensor product' of
nontrivial $H$-representations~\xcite{Warner:1983vz}.

The first detailed studies of these potentials were based on such
techniques, with which a number of solutions were found in the
80s. The present approach does not have such limitations and indeed
produced a large number of solutions that break $SO(8)$ completely and
leave no residual gauge symmetry. However, it still has
shortcomings. In particular, it randomly `fishes' for solutions and
will produce some stationary points with much greater likelihood than
others; in particular, relative probabilities are such that it seems
to effectively miss some solutions that were obtained by other means,
cf. the discussion of critical points {\bf S0983994} and {\bf
  S1400000}.

\section{Critical Points and their Properties}

The following tables list properties for all critical points that have
been found so far (either by conventional means, or by the novel
sensitivity backpropagation methods) and have a cosmological constant
in the range given in the title of this part of the series.

\smallbreak
\begin{longtable}{||l||}
\hline
\raise-0.1ex\hbox{
 \small {\bf S0600000}: \(V/g^2=-6.0\), Quality \(-\log_{10}(|Q|^2)=100.00\) , dim(GG)=28 \(\mathcal{N}=8\) }
\\
\hline
\hbox to6em{\(\phi\)}\quad\begin{minipage}{10.5cm}
{\vskip0.25ex
\small\tt
\(0\)
}\end{minipage}\\
\hline
\hbox to6em{\((m^2/m_0^2)[\psi]\)}\quad\begin{minipage}{8.5cm}
{\vskip0.25ex
\small
\(+1.000000\){\tiny \((\times 8)\)}
}\end{minipage}\\
\hline
\hbox to6em{\((m^2/m_0^2)[\chi]\)}\quad\begin{minipage}{8.5cm}
{\vskip0.25ex
\small
\(+0.000000\){\tiny \((\times 56)\)}
}\end{minipage}\\
\hline
\hbox to6em{\((m/m_0)[\phi]\)}\quad\begin{minipage}{8.5cm}
{\vskip0.25ex
\small
\(-2.000000\){\tiny \((\times 70)\)}
}\end{minipage}\\
\hline
\end{longtable}

\smallbreak\begin{longtable}{||l||}
\hline
\raise-0.1ex\hbox{
 \small {\bf S0668740}: \(V/g^2=-6.68740304976\), Quality \(-\log_{10}(|Q|^2)=22.46\) , dim(GG)=21}
\\
\hline
\hbox to6em{\(\phi\)}\quad\begin{minipage}{10.5cm}
{\vskip0.25ex
\small\tt
\(-0.201180_{1238s}\)  \(-0.201180_{1247s}\)  \(+0.201180_{1256s}\)  \(-0.201180_{1345s}\)  \(+0.201180_{1367s}\)  \(-0.201180_{1468s}\)  \(-0.201180_{1578s}\)  \(+0.201180_{2346s}\)  \(+0.201180_{2357s}\)  \(-0.201180_{2458s}\)  \(+0.201180_{2678s}\)  \(+0.201180_{3478s}\)  \(-0.201180_{3568s}\)  \(-0.201180_{4567s}\) 
}\end{minipage}\\
\hline
\hbox to6em{\((m^2/m_0^2)[\psi]\)}\quad\begin{minipage}{8.5cm}
{\vskip0.25ex
\small
\(+1.350000\){\tiny \((\times 8)\)}
}\end{minipage}\\
\hline
\hbox to6em{\((m^2/m_0^2)[\chi]\)}\quad\begin{minipage}{8.5cm}
{\vskip0.25ex
\small
\(+2.700000\){\tiny \((\times 8)\)}; \(+0.075000\){\tiny \((\times 48)\)}
}\end{minipage}\\
\hline
\hbox to6em{\((m/m_0)[\phi]\)}\quad\begin{minipage}{8.5cm}
{\vskip0.25ex
\small
\(+6.000000\); \(+0.000000\){\tiny \((\times 7)\)}; \(-1.200000\){\tiny \((\times 35)\)}; \(-2.400000\){\tiny \((\times 27)\)}
}\end{minipage}\\
\hline
\end{longtable}

\smallbreak\begin{longtable}{||l||}
\hline
\raise-0.1ex\hbox{
 \small {\bf S0698771}: \(V/g^2=-6.98771242969\), Quality \(-\log_{10}(|Q|^2)=30.91\) , dim(GG)=21}
\\
\hline
\hbox to6em{\(\phi\)}\quad\begin{minipage}{10.5cm}
{\vskip0.25ex
\small\tt
\(+0.240606_{1234c}\)  \(-0.240606_{1258c}\)  \(-0.240606_{1267c}\)  \(+0.240606_{1357c}\)  \(-0.240606_{1368c}\)  \(-0.240606_{1456c}\)  \(-0.240606_{1478c}\)  \(+0.240606_{2356c}\)  \(+0.240606_{2378c}\)  \(+0.240606_{2457c}\)  \(-0.240606_{2468c}\)  \(+0.240606_{3458c}\)  \(+0.240606_{3467c}\)  \(-0.240606_{5678c}\) 
}\end{minipage}\\
\hline
\hbox to6em{\((m^2/m_0^2)[\psi]\)}\quad\begin{minipage}{8.5cm}
{\vskip0.25ex
\small
\(+1.350000\){\tiny \((\times 8)\)}
}\end{minipage}\\
\hline
\hbox to6em{\((m^2/m_0^2)[\chi]\)}\quad\begin{minipage}{8.5cm}
{\vskip0.25ex
\small
\(+2.700000\){\tiny \((\times 8)\)}; \(+0.075000\){\tiny \((\times 48)\)}
}\end{minipage}\\
\hline
\hbox to6em{\((m/m_0)[\phi]\)}\quad\begin{minipage}{8.5cm}
{\vskip0.25ex
\small
\(+6.000000\); \(+0.000000\){\tiny \((\times 7)\)}; \(-1.200000\){\tiny \((\times 35)\)}; \(-2.400000\){\tiny \((\times 27)\)}
}\end{minipage}\\
\hline
\end{longtable}

\smallbreak\begin{longtable}{||l||}
\hline
\raise-0.1ex\hbox{
 \small {\bf S0719157}: \(V/g^2=-7.1915756261\), Quality \(-\log_{10}(|Q|^2)=31.17\) , dim(GG)=14 \(\mathcal{N}=1\) }
\\
\hline
\hbox to6em{\(\phi\)}\quad\begin{minipage}{10.5cm}
{\vskip0.25ex
\small\tt
\(-0.145735_{1234s}\)  \(-0.145735_{1257s}\)  \(-0.145735_{1268s}\)  \(+0.145735_{1358s}\)  \(-0.145735_{1367s}\)  \(+0.145735_{1456s}\)  \(-0.145735_{1478s}\)  \(-0.145735_{2356s}\)  \(+0.145735_{2378s}\)  \(+0.145735_{2458s}\)  \(-0.145735_{2467s}\)  \(+0.145735_{3457s}\)  \(+0.145735_{3468s}\)  \(-0.145735_{5678s}\)  \(+0.213923_{1234c}\)  \(-0.213923_{1257c}\)  \(-0.213923_{1268c}\)  \(-0.213923_{1356c}\)  \(+0.213923_{1378c}\)  \(-0.213923_{1458c}\)  \(+0.213923_{1467c}\)  \(-0.213923_{2358c}\)  \(+0.213923_{2367c}\)  \(+0.213923_{2456c}\)  \(-0.213923_{2478c}\)  \(-0.213923_{3457c}\)  \(-0.213923_{3468c}\)  \(-0.213923_{5678c}\) 
}\end{minipage}\\
\hline
\hbox to6em{\((m^2/m_0^2)[\psi]\)}\quad\begin{minipage}{8.5cm}
{\vskip0.25ex
\small
\(+1.500000\){\tiny \((\times 7)\)}; \(+1.000000\)
}\end{minipage}\\
\hline
\hbox to6em{\((m^2/m_0^2)[\chi]\)}\quad\begin{minipage}{8.5cm}
{\vskip0.25ex
\small
\(+3.000000\){\tiny \((\times 8)\)}; \(+0.750000\){\tiny \((\times 7)\)}; \(+0.083333\){\tiny \((\times 27)\)}; \(+0.000000\){\tiny \((\times 14)\)}
}\end{minipage}\\
\hline
\hbox to6em{\((m/m_0)[\phi]\)}\quad\begin{minipage}{8.5cm}
{\vskip0.25ex
\small
\(+6.449490\); \(+1.550510\); \(+0.000000\){\tiny \((\times 14)\)}; \(-1.425085\){\tiny \((\times 27)\)}; \(-2.241582\){\tiny \((\times 27)\)}
}\end{minipage}\\
\hline
\end{longtable}

\smallbreak\begin{longtable}{||l||}
\hline
\raise-0.1ex\hbox{
 \small {\bf S0779422}: \(V/g^2=-7.79422863406\), Quality \(-\log_{10}(|Q|^2)=22.80\) , dim(GG)=9 \(\mathcal{N}=2\) }
\\
\hline
\hbox to6em{\(\phi\)}\quad\begin{minipage}{10.5cm}
{\vskip0.25ex
\small\tt
\(+0.274653_{1238s}\)  \(-0.274653_{1346s}\)  \(+0.274653_{1357s}\)  \(+0.274653_{2468s}\)  \(-0.274653_{2578s}\)  \(+0.274653_{4567s}\)  \(-0.329239_{1247c}\)  \(-0.329239_{1256c}\)  \(+0.329239_{1345c}\)  \(+0.329239_{1367c}\)  \(+0.329239_{2458c}\)  \(+0.329239_{2678c}\)  \(+0.329239_{3478c}\)  \(+0.329239_{3568c}\) 
}\end{minipage}\\
\hline
\hbox to6em{\((m^2/m_0^2)[\psi]\)}\quad\begin{minipage}{8.5cm}
{\vskip0.25ex
\small
\(+1.777778\){\tiny \((\times 6)\)}; \(+1.000000\){\tiny \((\times 2)\)}
}\end{minipage}\\
\hline
\hbox to6em{\((m^2/m_0^2)[\chi]\)}\quad\begin{minipage}{8.5cm}
{\vskip0.25ex
\small
\(+3.555556\){\tiny \((\times 6)\)}; \(+3.280776\){\tiny \((\times 2)\)}; \(+1.219224\){\tiny \((\times 2)\)}; \(+0.888889\){\tiny \((\times 12)\)}; \(+0.055556\){\tiny \((\times 18)\)}; \(+0.000000\){\tiny \((\times 16)\)}
}\end{minipage}\\
\hline
\hbox to6em{\((m/m_0)[\phi]\)}\quad\begin{minipage}{8.5cm}
{\vskip0.25ex
\small
\(+7.123106\); \(+2.000000\){\tiny \((\times 3)\)}; \(+0.000000\){\tiny \((\times 19)\)}; \(-1.123106\); \(-1.555556\){\tiny \((\times 18)\)}; \(-2.000000\){\tiny \((\times 16)\)}; \(-2.222222\){\tiny \((\times 12)\)}
}\end{minipage}\\
\hline
\end{longtable}

\smallbreak\begin{longtable}{||l||}
\hline
\raise-0.1ex\hbox{
 \small {\bf S0800000}: \(V/g^2=-8.0\), Quality \(-\log_{10}(|Q|^2)=107.22\) , dim(GG)=15}
\\
\hline
\hbox to6em{\(\phi\)}\quad\begin{minipage}{10.5cm}
{\vskip0.25ex
\small\tt
\(+0.440687_{1357c}\)  \(-0.440687_{1368c}\)  \(-0.440687_{1458c}\)  \(-0.440687_{1467c}\)  \(+0.440687_{2358c}\)  \(+0.440687_{2367c}\)  \(+0.440687_{2457c}\)  \(-0.440687_{2468c}\) 
}\end{minipage}\\
\hline
\hbox to6em{\((m^2/m_0^2)[\psi]\)}\quad\begin{minipage}{8.5cm}
{\vskip0.25ex
\small
\(+1.687500\){\tiny \((\times 8)\)}
}\end{minipage}\\
\hline
\hbox to6em{\((m^2/m_0^2)[\chi]\)}\quad\begin{minipage}{8.5cm}
{\vskip0.25ex
\small
\(+3.375000\){\tiny \((\times 8)\)}; \(+2.343750\){\tiny \((\times 8)\)}; \(+0.093750\){\tiny \((\times 40)\)}
}\end{minipage}\\
\hline
\hbox to6em{\((m/m_0)[\phi]\)}\quad\begin{minipage}{8.5cm}
{\vskip0.25ex
\small
\(+6.000000\){\tiny \((\times 2)\)}; \(+0.000000\){\tiny \((\times 28)\)}; \(-0.750000\){\tiny \((\times 20)\)}; \(-3.000000\){\tiny \((\times 20)\)}
}\end{minipage}\\
\hline
\end{longtable}

\smallbreak\begin{longtable}{||l||}
\hline
\raise-0.1ex\hbox{
 \small {\bf S0847213}: \(V/g^2=-8.472135955\), Quality \(-\log_{10}(|Q|^2)=107.66\) , dim(GG)=5}
\\
\hline
\hbox to6em{\(\phi\)}\quad\begin{minipage}{10.5cm}
{\vskip0.25ex
\small\tt
\(+0.421241_{1236s}\)  \(+0.402359_{1456s}\)  \(+0.421241_{1678s}\)  \(+0.421241_{2345s}\)  \(+0.402359_{2378s}\)  \(+0.421241_{4578s}\)  \(+0.421241_{1268c}\)  \(+0.421241_{1367c}\)  \(+0.421241_{2458c}\)  \(+0.421241_{3457c}\) 
}\end{minipage}\\
\hline
\hbox to6em{\((m^2/m_0^2)[\psi]\)}\quad\begin{minipage}{8.5cm}
{\vskip0.25ex
\small
\(+2.170820\){\tiny \((\times 4)\)}; \(+1.570820\){\tiny \((\times 4)\)}
}\end{minipage}\\
\hline
\hbox to6em{\((m^2/m_0^2)[\chi]\)}\quad\begin{minipage}{8.5cm}
{\vskip0.25ex
\small
\(+4.341641\){\tiny \((\times 4)\)}; \(+3.645639\){\tiny \((\times 4)\)}; \(+3.141641\){\tiny \((\times 4)\)}; \(+2.206517\){\tiny \((\times 4)\)}; \(+0.556231\){\tiny \((\times 8)\)}; \(+0.414590\){\tiny \((\times 4)\)}; \(+0.291745\){\tiny \((\times 4)\)}; \(+0.114590\){\tiny \((\times 16)\)}; \(+0.039070\){\tiny \((\times 4)\)}; \(+0.023104\){\tiny \((\times 4)\)}
}\end{minipage}\\
\hline
\hbox to6em{\((m/m_0)[\phi]\)}\quad\begin{minipage}{8.5cm}
{\vskip0.25ex
\small
\(+7.295813\){\tiny \((\times 2)\)}; \(+7.120248\); \(+2.966563\){\tiny \((\times 2)\)}; \(+0.283282\){\tiny \((\times 2)\)}; \(+0.000000\){\tiny \((\times 23)\)}; \(-0.633437\); \(-1.266874\){\tiny \((\times 6)\)}; \(-1.393775\){\tiny \((\times 8)\)}; \(-1.416408\){\tiny \((\times 6)\)}; \(-1.779716\){\tiny \((\times 2)\)}; \(-2.098021\){\tiny \((\times 8)\)}; \(-2.387121\); \(-2.400000\){\tiny \((\times 6)\)}; \(-2.832816\){\tiny \((\times 2)\)}
}\end{minipage}\\
\hline
\end{longtable}

\smallbreak\begin{longtable}{||l||}
\hline
\raise-0.1ex\hbox{
 \small {\bf S0869596}: \(V/g^2=-8.69596907922\), Quality \(-\log_{10}(|Q|^2)=107.59\) , dim(GG)=4}
\\
\hline
\hbox to6em{\(\phi\)}\quad\begin{minipage}{10.5cm}
{\vskip0.25ex
\small\tt
\(+0.450223_{1237s}\)  \(-0.450223_{1246s}\)  \(-0.450223_{1258s}\)  \(-0.450223_{3467s}\)  \(+0.450223_{3578s}\)  \(-0.450223_{4568s}\)  \(-0.167520_{1345c}\)  \(+0.167520_{1367c}\)  \(+0.502560_{1478c}\)  \(+0.167520_{1568c}\)  \(-0.167520_{2347c}\)  \(-0.502560_{2356c}\)  \(+0.167520_{2458c}\)  \(-0.167520_{2678c}\) 
}\end{minipage}\\
\hline
\hbox to6em{\((m^2/m_0^2)[\psi]\)}\quad\begin{minipage}{8.5cm}
{\vskip0.25ex
\small
\(+2.326179\){\tiny \((\times 2)\)}; \(+1.874815\){\tiny \((\times 6)\)}
}\end{minipage}\\
\hline
\hbox to6em{\((m^2/m_0^2)[\chi]\)}\quad\begin{minipage}{8.5cm}
{\vskip0.25ex
\small
\(+4.652358\){\tiny \((\times 2)\)}; \(+3.749630\){\tiny \((\times 6)\)}; \(+3.735622\){\tiny \((\times 6)\)}; \(+2.723885\){\tiny \((\times 2)\)}; \(+0.515437\){\tiny \((\times 10)\)}; \(+0.369264\){\tiny \((\times 10)\)}; \(+0.278072\){\tiny \((\times 6)\)}; \(+0.124339\){\tiny \((\times 6)\)}; \(+0.081783\){\tiny \((\times 2)\)}; \(+0.003699\){\tiny \((\times 6)\)}
}\end{minipage}\\
\hline
\hbox to6em{\((m/m_0)[\phi]\)}\quad\begin{minipage}{8.5cm}
{\vskip0.25ex
\small
\(+9.302691\){\tiny \((\times 2)\)}; \(+6.997127\); \(+1.964690\){\tiny \((\times 5)\)}; \(+0.000000\){\tiny \((\times 24)\)}; \(-0.295834\){\tiny \((\times 5)\)}; \(-0.878409\){\tiny \((\times 2)\)}; \(-1.067879\); \(-1.284943\){\tiny \((\times 6)\)}; \(-1.319396\){\tiny \((\times 2)\)}; \(-1.718333\){\tiny \((\times 3)\)}; \(-2.208712\){\tiny \((\times 3)\)}; \(-2.260625\){\tiny \((\times 10)\)}; \(-2.507722\){\tiny \((\times 5)\)}; \(-3.230157\)
}\end{minipage}\\
\hline
\end{longtable}

\smallbreak\begin{longtable}{||l||}
\hline
\raise-0.1ex\hbox{
 \small {\bf S0880733}: \(V/g^2=-8.80733895008\), Quality \(-\log_{10}(|Q|^2)=33.55\) , dim(GG)=6}
\\
\hline
\hbox to6em{\(\phi\)}\quad\begin{minipage}{10.5cm}
{\vskip0.25ex
\small\tt
\(+0.466566_{1278s}\)  \(+0.466566_{1368s}\)  \(+0.466566_{1458s}\)  \(+0.466566_{2367s}\)  \(+0.466566_{2457s}\)  \(+0.466566_{3456s}\)  \(-0.286554_{1235c}\)  \(-0.286554_{1268c}\)  \(+0.286554_{1378c}\)  \(+0.286554_{1567c}\)  \(+0.286554_{2348c}\)  \(+0.286554_{2456c}\)  \(-0.286554_{3457c}\)  \(-0.286554_{4678c}\) 
}\end{minipage}\\
\hline
\hbox to6em{\((m^2/m_0^2)[\psi]\)}\quad\begin{minipage}{8.5cm}
{\vskip0.25ex
\small
\(+2.049038\){\tiny \((\times 8)\)}
}\end{minipage}\\
\hline
\hbox to6em{\((m^2/m_0^2)[\chi]\)}\quad\begin{minipage}{8.5cm}
{\vskip0.25ex
\small
\(+4.098076\){\tiny \((\times 8)\)}; \(+3.774434\){\tiny \((\times 8)\)}; \(+0.390544\){\tiny \((\times 32)\)}; \(+0.006655\){\tiny \((\times 8)\)}
}\end{minipage}\\
\hline
\hbox to6em{\((m/m_0)[\phi]\)}\quad\begin{minipage}{8.5cm}
{\vskip0.25ex
\small
\(+10.596546\); \(+10.098076\); \(+6.928203\); \(+1.464102\){\tiny \((\times 9)\)}; \(+0.000000\){\tiny \((\times 22)\)}; \(-0.535898\){\tiny \((\times 9)\)}; \(-1.132444\); \(-1.901924\){\tiny \((\times 15)\)}; \(-2.535898\){\tiny \((\times 10)\)}; \(-3.464102\)
}\end{minipage}\\
\hline
\end{longtable}

\smallbreak\begin{longtable}{||l||}
\hline
\raise-0.1ex\hbox{
 \small {\bf S0983994}: \(V/g^2=-9.83994835633\), Quality \(-\log_{10}(|Q|^2)=32.75\) , dim(GG)=4}
\\
\hline
\hbox to6em{\(\phi\)}\quad\begin{minipage}{10.5cm}
{\vskip0.25ex
\small\tt
\(+0.186375_{1235s}\)  \(+0.186375_{1246s}\)  \(+0.145475_{1278s}\)  \(-0.186375_{1347s}\)  \(+0.227274_{1368s}\)  \(+0.518225_{1458s}\)  \(+0.186375_{1567s}\)  \(-0.186375_{2348s}\)  \(+0.518225_{2367s}\)  \(+0.227274_{2457s}\)  \(+0.186375_{2568s}\)  \(+0.145475_{3456s}\)  \(-0.186375_{3578s}\)  \(-0.186375_{4678s}\)  \(+0.510767_{1278c}\)  \(+0.510767_{1367c}\)  \(-0.510767_{1457c}\)  \(-0.510767_{2368c}\)  \(+0.510767_{2458c}\)  \(-0.510767_{3456c}\) 
}\end{minipage}\\
\hline
\hbox to6em{\((m^2/m_0^2)[\psi]\)}\quad\begin{minipage}{8.5cm}
{\vskip0.25ex
\small
\(+2.703125\){\tiny \((\times 2)\)}; \(+1.953125\){\tiny \((\times 6)\)}
}\end{minipage}\\
\hline
\hbox to6em{\((m^2/m_0^2)[\chi]\)}\quad\begin{minipage}{8.5cm}
{\vskip0.25ex
\small
\(+5.406250\){\tiny \((\times 2)\)}; \(+3.962327\){\tiny \((\times 6)\)}; \(+3.906250\){\tiny \((\times 6)\)}; \(+2.039063\){\tiny \((\times 2)\)}; \(+1.549016\){\tiny \((\times 10)\)}; \(+0.278228\){\tiny \((\times 6)\)}; \(+0.164063\){\tiny \((\times 2)\)}; \(+0.164062\){\tiny \((\times 4)\)}; \(+0.139132\){\tiny \((\times 6)\)}; \(+0.126563\){\tiny \((\times 2)\)}; \(+0.079109\){\tiny \((\times 10)\)}
}\end{minipage}\\
\hline
\hbox to6em{\((m/m_0)[\phi]\)}\quad\begin{minipage}{8.5cm}
{\vskip0.25ex
\small
\(+8.148231\){\tiny \((\times 2)\)}; \(+7.349345\); \(+4.717019\){\tiny \((\times 5)\)}; \(+0.000000\){\tiny \((\times 24)\)}; \(-0.993334\); \(-1.000000\){\tiny \((\times 3)\)}; \(-1.123231\){\tiny \((\times 2)\)}; \(-1.387500\){\tiny \((\times 10)\)}; \(-1.405110\){\tiny \((\times 5)\)}; \(-1.487500\){\tiny \((\times 6)\)}; \(-1.750000\){\tiny \((\times 2)\)}; \(-1.911909\){\tiny \((\times 5)\)}; \(-2.400000\){\tiny \((\times 3)\)}; \(-3.156012\)
}\end{minipage}\\
\hline
\end{longtable}

\smallbreak\begin{longtable}{||l||}
\hline
\raise-0.1ex\hbox{
 \small {\bf S0998708}: \(V/g^2=-9.98708340034\), Quality \(-\log_{10}(|Q|^2)=34.85\) , \(U(1)\) }
\\
\hline
\hbox to6em{\(\phi\)}\quad\begin{minipage}{10.5cm}
{\vskip0.25ex
\small\tt
\(+0.004542_{1236s}\)  \(+0.132440_{1237s}\)  \(+0.106833_{1238s}\)  \(+0.162666_{1247s}\)  \(-0.121240_{1248s}\)  \(+0.523185_{1256s}\)  \(-0.100317_{1258s}\)  \(+0.179186_{1345s}\)  \(-0.167907_{1367s}\)  \(+0.132440_{1368s}\)  \(+0.004542_{1378s}\)  \(+0.121240_{1467s}\)  \(+0.155365_{1468s}\)  \(-0.100317_{1567s}\)  \(+0.172114_{1578s}\)  \(-0.172114_{2346s}\)  \(+0.100317_{2348s}\)  \(-0.155365_{2357s}\)  \(+0.121240_{2358s}\)  \(-0.004542_{2456s}\)  \(+0.132440_{2457s}\)  \(+0.167907_{2458s}\)  \(-0.179186_{2678s}\)  \(-0.100317_{3467s}\)  \(+0.523185_{3478s}\)  \(+0.121240_{3567s}\)  \(+0.162666_{3568s}\)  \(+0.106833_{4567s}\)  \(-0.132440_{4568s}\)  \(+0.004542_{4578s}\)  \(-0.102479_{1236c}\)  \(+0.004626_{1237c}\)  \(+0.541792_{1238c}\)  \(+0.065978_{1246c}\)  \(+0.032183_{1248c}\)  \(-0.093253_{1257c}\)  \(+0.007467_{1258c}\)  \(+0.111285_{1345c}\)  \(+0.078664_{1367c}\)  \(+0.004626_{1368c}\)  \(-0.102479_{1378c}\)  \(+0.032183_{1467c}\)  \(-0.363013_{1478c}\)  \(-0.007467_{1567c}\)  \(+0.558308_{1568c}\)  \(-0.558308_{2347c}\)  \(-0.007467_{2348c}\)  \(+0.363013_{2356c}\)  \(-0.032183_{2358c}\)  \(-0.102479_{2456c}\)  \(-0.004626_{2457c}\)  \(+0.078664_{2458c}\)  \(+0.111285_{2678c}\)  \(-0.007467_{3467c}\)  \(-0.093253_{3468c}\)  \(+0.032183_{3567c}\)  \(+0.065978_{3578c}\)  \(-0.541792_{4567c}\)  \(+0.004626_{4568c}\)  \(+0.102479_{4578c}\) 
}\end{minipage}\\
\hline
\hbox to6em{\((m^2/m_0^2)[\psi]\)}\quad\begin{minipage}{8.5cm}
{\vskip0.25ex
\small
\(+2.797126\){\tiny \((\times 2)\)}; \(+2.197173\){\tiny \((\times 2)\)}; \(+2.080922\); \(+1.890465\){\tiny \((\times 2)\)}; \(+1.597929\)
}\end{minipage}\\
\hline
\end{longtable}

\smallbreak\begin{longtable}{||l||}
\hline
\raise-0.1ex\hbox{
 \small {\bf S1006758}: \(V/g^2=-10.0675803596\), Quality \(-\log_{10}(|Q|^2)=30.19\) , \(U(1)\) }
\\
\hline
\hbox to6em{\(\phi\)}\quad\begin{minipage}{10.5cm}
{\vskip0.25ex
\small\tt
\(+0.543353_{1236s}\)  \(-0.012650_{1245s}\)  \(+0.185896_{1247s}\)  \(-0.178360_{1258s}\)  \(-0.024558_{1278s}\)  \(-0.167188_{1348s}\)  \(-0.234276_{1357s}\)  \(-0.185896_{1456s}\)  \(-0.012650_{1467s}\)  \(+0.024558_{1568s}\)  \(-0.178360_{1678s}\)  \(-0.178360_{2345s}\)  \(+0.024558_{2347s}\)  \(-0.012650_{2358s}\)  \(-0.185896_{2378s}\)  \(-0.234276_{2468s}\)  \(-0.167188_{2567s}\)  \(-0.024558_{3456s}\)  \(-0.178360_{3467s}\)  \(+0.185896_{3568s}\)  \(-0.012650_{3678s}\)  \(+0.543353_{4578s}\)  \(+0.342403_{1235c}\)  \(-0.113920_{1246c}\)  \(+0.025164_{1247c}\)  \(-0.005184_{1268c}\)  \(+0.139237_{1278c}\)  \(-0.565212_{1348c}\)  \(+0.560180_{1367c}\)  \(+0.025164_{1456c}\)  \(-0.113920_{1457c}\)  \(+0.139237_{1568c}\)  \(-0.005184_{1578c}\)  \(-0.005184_{2346c}\)  \(-0.139237_{2347c}\)  \(-0.113920_{2368c}\)  \(-0.025164_{2378c}\)  \(+0.560180_{2458c}\)  \(+0.565212_{2567c}\)  \(-0.139237_{3456c}\)  \(-0.005184_{3457c}\)  \(-0.025164_{3568c}\)  \(-0.113920_{3578c}\)  \(+0.342403_{4678c}\) 
}\end{minipage}\\
\hline
\hbox to6em{\((m^2/m_0^2)[\psi]\)}\quad\begin{minipage}{8.5cm}
{\vskip0.25ex
\small
\(+2.963207\); \(+2.470417\); \(+2.223788\); \(+2.215048\){\tiny \((\times 2)\)}; \(+2.033524\); \(+1.779289\){\tiny \((\times 2)\)}
}\end{minipage}\\
\hline
\end{longtable}

\smallbreak\begin{longtable}{||l||}
\hline
\raise-0.1ex\hbox{
 \small {\bf S1039624}: \(V/g^2=-10.396246071\), Quality \(-\log_{10}(|Q|^2)=33.84\) , \(U(1)\) }
\\
\hline
\hbox to6em{\(\phi\)}\quad\begin{minipage}{10.5cm}
{\vskip0.25ex
\small\tt
\(-0.105452_{1234s}\)  \(+0.117214_{1235s}\)  \(+0.189713_{1238s}\)  \(+0.117214_{1246s}\)  \(-0.189713_{1247s}\)  \(-0.561575_{1256s}\)  \(+0.062221_{1278s}\)  \(-0.189713_{1346s}\)  \(+0.117214_{1347s}\)  \(-0.161612_{1357s}\)  \(+0.052498_{1368s}\)  \(-0.204843_{1458s}\)  \(-0.028813_{1467s}\)  \(-0.117214_{1567s}\)  \(+0.189713_{1678s}\)  \(+0.189713_{2345s}\)  \(+0.117214_{2348s}\)  \(-0.028813_{2358s}\)  \(-0.204843_{2367s}\)  \(+0.052498_{2457s}\)  \(-0.161612_{2468s}\)  \(-0.117214_{2568s}\)  \(-0.189713_{2578s}\)  \(+0.062221_{3456s}\)  \(-0.561575_{3478s}\)  \(-0.189713_{3568s}\)  \(-0.117214_{3578s}\)  \(+0.189713_{4567s}\)  \(-0.117214_{4678s}\)  \(-0.105452_{5678s}\)  \(+0.092339_{1235c}\)  \(+0.005507_{1238c}\)  \(-0.092339_{1246c}\)  \(-0.005507_{1247c}\)  \(+0.633469_{1278c}\)  \(+0.005507_{1346c}\)  \(+0.092339_{1347c}\)  \(+0.606052_{1357c}\)  \(-0.092339_{1567c}\)  \(+0.005507_{1678c}\)  \(-0.005507_{2345c}\)  \(-0.092339_{2348c}\)  \(-0.606052_{2468c}\)  \(+0.092339_{2568c}\)  \(-0.005507_{2578c}\)  \(-0.633469_{3456c}\)  \(+0.005507_{3568c}\)  \(-0.092339_{3578c}\)  \(-0.005507_{4567c}\)  \(+0.092339_{4678c}\) 
}\end{minipage}\\
\hline
\hbox to6em{\((m^2/m_0^2)[\psi]\)}\quad\begin{minipage}{8.5cm}
{\vskip0.25ex
\small
\(+2.956723\){\tiny \((\times 2)\)}; \(+2.289515\){\tiny \((\times 2)\)}; \(+2.230739\){\tiny \((\times 2)\)}; \(+1.656801\){\tiny \((\times 2)\)}
}\end{minipage}\\
\hline
\end{longtable}

\smallbreak\begin{longtable}{||l||}
\hline
\raise-0.1ex\hbox{
 \small {\bf S1043471}: \(V/g^2=-10.434712595\), Quality \(-\log_{10}(|Q|^2)=34.40\) }
\\
\hline
\hbox to6em{\(\phi\)}\quad\begin{minipage}{10.5cm}
{\vskip0.25ex
\small\tt
\(+0.000744_{1235s}\)  \(+0.218613_{1236s}\)  \(+0.014376_{1237s}\)  \(+0.014376_{1246s}\)  \(-0.600698_{1247s}\)  \(+0.014376_{1248s}\)  \(+0.014376_{1257s}\)  \(+0.218613_{1258s}\)  \(+0.000744_{1268s}\)  \(-0.000744_{1346s}\)  \(-0.191674_{1348s}\)  \(+0.191674_{1357s}\)  \(-0.014376_{1368s}\)  \(-0.010668_{1378s}\)  \(-0.191674_{1456s}\)  \(-0.000744_{1458s}\)  \(-0.014376_{1467s}\)  \(+0.014376_{1478s}\)  \(+0.010668_{1567s}\)  \(+0.014376_{1568s}\)  \(+0.191674_{1678s}\)  \(+0.191674_{2345s}\)  \(+0.014376_{2347s}\)  \(-0.010668_{2348s}\)  \(+0.014376_{2356s}\)  \(-0.014376_{2358s}\)  \(-0.000744_{2367s}\)  \(-0.191674_{2378s}\)  \(+0.010668_{2456s}\)  \(-0.014376_{2457s}\)  \(+0.191674_{2468s}\)  \(-0.191674_{2567s}\)  \(-0.000744_{2578s}\)  \(-0.000744_{3457s}\)  \(+0.218613_{3467s}\)  \(-0.014376_{3468s}\)  \(-0.014376_{3567s}\)  \(-0.600698_{3568s}\)  \(-0.014376_{3578s}\)  \(-0.014376_{4568s}\)  \(+0.218613_{4578s}\)  \(-0.000744_{4678s}\)  \(+0.068273_{1237c}\)  \(-0.278841_{1238c}\)  \(+0.068273_{1246c}\)  \(-0.038078_{1247c}\)  \(+0.068273_{1248c}\)  \(-0.306271_{1256c}\)  \(+0.068273_{1257c}\)  \(+0.655534_{1345c}\)  \(+0.306271_{1367c}\)  \(+0.068273_{1368c}\)  \(-0.068273_{1467c}\)  \(+0.068273_{1478c}\)  \(-0.068273_{1568c}\)  \(-0.278841_{1578c}\)  \(-0.278841_{2346c}\)  \(+0.068273_{2347c}\)  \(-0.068273_{2356c}\)  \(+0.068273_{2358c}\)  \(-0.068273_{2457c}\)  \(+0.306271_{2458c}\)  \(+0.655534_{2678c}\)  \(+0.068273_{3468c}\)  \(+0.306271_{3478c}\)  \(+0.068273_{3567c}\)  \(+0.038078_{3568c}\)  \(+0.068273_{3578c}\)  \(+0.278841_{4567c}\)  \(+0.068273_{4568c}\) 
}\end{minipage}\\
\hline
\end{longtable}

\smallbreak\begin{longtable}{||l||}
\hline
\raise-0.1ex\hbox{
 \small {\bf S1046017}: \(V/g^2=-10.4601767468\), Quality \(-\log_{10}(|Q|^2)=21.53\) }
\\
\hline
\hbox to6em{\(\phi\)}\quad\begin{minipage}{10.5cm}
{\vskip0.25ex
\small\tt
\(+0.611785_{1238s}\)  \(+0.179220_{1245s}\)  \(+0.186916_{1267s}\)  \(+0.179220_{1347s}\)  \(+0.186916_{1356s}\)  \(-0.214745_{1468s}\)  \(+0.237954_{1578s}\)  \(-0.237954_{2346s}\)  \(+0.214745_{2357s}\)  \(-0.186916_{2478s}\)  \(-0.179220_{2568s}\)  \(+0.186916_{3458s}\)  \(+0.179220_{3678s}\)  \(+0.611785_{4567s}\)  \(+0.335852_{1258c}\)  \(+0.335852_{1267c}\)  \(+0.335852_{1356c}\)  \(-0.335852_{1378c}\)  \(+0.603656_{1468c}\)  \(+0.603656_{2357c}\)  \(-0.335852_{2456c}\)  \(+0.335852_{2478c}\)  \(-0.335852_{3458c}\)  \(-0.335852_{3467c}\) 
}\end{minipage}\\
\hline
\hbox to6em{\((m^2/m_0^2)[\psi]\)}\quad\begin{minipage}{8.5cm}
{\vskip0.25ex
\small
\(+2.664733\){\tiny \((\times 4)\)}; \(+1.954969\){\tiny \((\times 4)\)}
}\end{minipage}\\
\hline
\hbox to6em{\((m^2/m_0^2)[\chi]\)}\quad\begin{minipage}{8.5cm}
{\vskip0.25ex
\small
\(+5.329467\){\tiny \((\times 4)\)}; \(+3.909937\){\tiny \((\times 4)\)}; \(+3.845437\){\tiny \((\times 4)\)}; \(+3.728603\){\tiny \((\times 4)\)}; \(+2.650581\){\tiny \((\times 4)\)}; \(+1.395470\){\tiny \((\times 4)\)}; \(+1.074521\){\tiny \((\times 4)\)}; \(+0.806297\){\tiny \((\times 4)\)}; \(+0.508608\){\tiny \((\times 4)\)}; \(+0.123810\){\tiny \((\times 4)\)}; \(+0.092082\){\tiny \((\times 4)\)}; \(+0.087909\){\tiny \((\times 4)\)}; \(+0.014791\){\tiny \((\times 4)\)}; \(+0.009805\){\tiny \((\times 4)\)}
}\end{minipage}\\
\hline
\end{longtable}

\smallbreak\begin{longtable}{||l||}
\hline
\raise-0.1ex\hbox{
 \small {\bf S1067475}: \(V/g^2=-10.674754183\), Quality \(-\log_{10}(|Q|^2)=22.69\) , \(U(1)\times U(1)\) }
\\
\hline
\hbox to6em{\(\phi\)}\quad\begin{minipage}{10.5cm}
{\vskip0.25ex
\small\tt
\(-0.593541_{1235s}\)  \(+0.033813_{1246s}\)  \(-0.199941_{1248s}\)  \(-0.199941_{1267s}\)  \(+0.031708_{1278s}\)  \(-0.199941_{1346s}\)  \(+0.031708_{1348s}\)  \(+0.033813_{1367s}\)  \(-0.199941_{1378s}\)  \(+0.269637_{1457s}\)  \(+0.195765_{1568s}\)  \(+0.195765_{2347s}\)  \(-0.269637_{2368s}\)  \(+0.199941_{2456s}\)  \(-0.033813_{2458s}\)  \(+0.031708_{2567s}\)  \(-0.199941_{2578s}\)  \(+0.031708_{3456s}\)  \(-0.199941_{3458s}\)  \(+0.199941_{3567s}\)  \(-0.033813_{3578s}\)  \(+0.593541_{4678s}\)  \(+0.328895_{1236c}\)  \(-0.328895_{1245c}\)  \(-0.328895_{1357c}\)  \(+0.328895_{1467c}\)  \(+0.657790_{1568c}\)  \(-0.657790_{2347c}\)  \(-0.328895_{2358c}\)  \(+0.328895_{2468c}\)  \(+0.328895_{3678c}\)  \(-0.328895_{4578c}\) 
}\end{minipage}\\
\hline
\hbox to6em{\((m^2/m_0^2)[\psi]\)}\quad\begin{minipage}{8.5cm}
{\vskip0.25ex
\small
\(+2.655755\){\tiny \((\times 4)\)}; \(+2.136690\){\tiny \((\times 4)\)}
}\end{minipage}\\
\hline
\hbox to6em{\((m^2/m_0^2)[\chi]\)}\quad\begin{minipage}{8.5cm}
{\vskip0.25ex
\small
\(+5.311511\){\tiny \((\times 4)\)}; \(+4.273380\){\tiny \((\times 4)\)}; \(+3.994898\){\tiny \((\times 4)\)}; \(+3.672547\){\tiny \((\times 4)\)}; \(+3.462577\){\tiny \((\times 4)\)}; \(+1.344334\){\tiny \((\times 4)\)}; \(+1.194729\){\tiny \((\times 4)\)}; \(+0.802447\){\tiny \((\times 4)\)}; \(+0.637759\){\tiny \((\times 4)\)}; \(+0.177085\){\tiny \((\times 4)\)}; \(+0.090020\){\tiny \((\times 4)\)}; \(+0.085628\){\tiny \((\times 4)\)}; \(+0.084559\){\tiny \((\times 4)\)}; \(+0.000532\){\tiny \((\times 4)\)}
}\end{minipage}\\
\hline
\end{longtable}

\smallbreak\begin{longtable}{||l||}
\hline
\raise-0.1ex\hbox{
 \small {\bf S1068971}: \(V/g^2=-10.6897177038\), Quality \(-\log_{10}(|Q|^2)=21.93\) , \(U(1)\times U(1)\) }
\\
\hline
\hbox to6em{\(\phi\)}\quad\begin{minipage}{10.5cm}
{\vskip0.25ex
\small\tt
\(-0.679593_{1345s}\)  \(+0.679593_{2678s}\)  \(+0.129174_{1235c}\)  \(-0.639602_{1246c}\)  \(+0.129174_{1278c}\)  \(+0.129174_{1348c}\)  \(+0.390696_{1367c}\)  \(-0.129174_{1457c}\)  \(+0.639602_{1568c}\)  \(-0.639602_{2347c}\)  \(-0.129174_{2368c}\)  \(+0.390696_{2458c}\)  \(-0.129174_{2567c}\)  \(-0.129174_{3456c}\)  \(-0.639602_{3578c}\)  \(+0.129174_{4678c}\) 
}\end{minipage}\\
\hline
\hbox to6em{\((m^2/m_0^2)[\psi]\)}\quad\begin{minipage}{8.5cm}
{\vskip0.25ex
\small
\(+2.597091\){\tiny \((\times 4)\)}; \(+2.039597\){\tiny \((\times 4)\)}
}\end{minipage}\\
\hline
\hbox to6em{\((m^2/m_0^2)[\chi]\)}\quad\begin{minipage}{8.5cm}
{\vskip0.25ex
\small
\(+5.194182\){\tiny \((\times 4)\)}; \(+4.079194\){\tiny \((\times 4)\)}; \(+4.026394\){\tiny \((\times 4)\)}; \(+3.908820\){\tiny \((\times 4)\)}; \(+2.484867\){\tiny \((\times 4)\)}; \(+1.127723\){\tiny \((\times 4)\)}; \(+0.910972\){\tiny \((\times 8)\)}; \(+0.786650\){\tiny \((\times 4)\)}; \(+0.107702\){\tiny \((\times 4)\)}; \(+0.052989\){\tiny \((\times 4)\)}; \(+0.049861\){\tiny \((\times 4)\)}; \(+0.044933\){\tiny \((\times 8)\)}
}\end{minipage}\\
\hline
\hbox to6em{\((m/m_0)[\phi]\)}\quad\begin{minipage}{8.5cm}
{\vskip0.25ex
\small
\(+9.851609\); \(+9.527399\); \(+7.238720\); \(+4.688077\); \(+4.120364\){\tiny \((\times 4)\)}; \(+2.824418\){\tiny \((\times 4)\)}; \(+2.634825\); \(+0.000000\){\tiny \((\times 26)\)}; \(-0.410361\); \(-1.033164\){\tiny \((\times 4)\)}; \(-1.167795\){\tiny \((\times 4)\)}; \(-1.357886\){\tiny \((\times 2)\)}; \(-1.493136\); \(-1.603920\){\tiny \((\times 4)\)}; \(-1.649304\); \(-1.915972\){\tiny \((\times 4)\)}; \(-2.169029\); \(-2.183838\){\tiny \((\times 4)\)}; \(-2.724879\){\tiny \((\times 4)\)}; \(-3.109369\)
}\end{minipage}\\
\hline
\end{longtable}

\smallbreak\begin{longtable}{||l||}
\hline
\raise-0.1ex\hbox{
 \small {\bf S1075828}: \(V/g^2=-10.7582870728\), Quality \(-\log_{10}(|Q|^2)=33.23\) , dim(GG)=4}
\\
\hline
\hbox to6em{\(\phi\)}\quad\begin{minipage}{10.5cm}
{\vskip0.25ex
\small\tt
\(+0.677442_{1238s}\)  \(+0.677442_{4567s}\)  \(-0.123683_{1238c}\)  \(+0.117884_{1247c}\)  \(-0.654358_{1256c}\)  \(-0.163725_{1345c}\)  \(-0.372749_{1367c}\)  \(+0.654358_{1468c}\)  \(+0.117884_{1578c}\)  \(+0.117884_{2346c}\)  \(+0.654358_{2357c}\)  \(-0.372749_{2458c}\)  \(-0.163725_{2678c}\)  \(+0.654358_{3478c}\)  \(-0.117884_{3568c}\)  \(+0.123683_{4567c}\) 
}\end{minipage}\\
\hline
\hbox to6em{\((m^2/m_0^2)[\psi]\)}\quad\begin{minipage}{8.5cm}
{\vskip0.25ex
\small
\(+2.562132\){\tiny \((\times 4)\)}; \(+2.137868\){\tiny \((\times 4)\)}
}\end{minipage}\\
\hline
\hbox to6em{\((m^2/m_0^2)[\chi]\)}\quad\begin{minipage}{8.5cm}
{\vskip0.25ex
\small
\(+5.124264\){\tiny \((\times 4)\)}; \(+4.275736\){\tiny \((\times 4)\)}; \(+3.987684\){\tiny \((\times 4)\)}; \(+3.982994\){\tiny \((\times 4)\)}; \(+2.787320\){\tiny \((\times 4)\)}; \(+1.105330\){\tiny \((\times 8)\)}; \(+0.879584\){\tiny \((\times 4)\)}; \(+0.693198\){\tiny \((\times 4)\)}; \(+0.107828\){\tiny \((\times 4)\)}; \(+0.104591\){\tiny \((\times 4)\)}; \(+0.056802\){\tiny \((\times 4)\)}; \(+0.044670\){\tiny \((\times 8)\)}
}\end{minipage}\\
\hline
\hbox to6em{\((m/m_0)[\phi]\)}\quad\begin{minipage}{8.5cm}
{\vskip0.25ex
\small
\(+10.094811\); \(+9.396152\); \(+7.145562\); \(+5.246738\){\tiny \((\times 3)\)}; \(+2.989975\){\tiny \((\times 6)\)}; \(+2.945602\); \(+0.439230\){\tiny \((\times 3)\)}; \(+0.000000\){\tiny \((\times 24)\)}; \(-0.800000\){\tiny \((\times 2)\)}; \(-0.989975\){\tiny \((\times 6)\)}; \(-0.996152\); \(-1.272303\); \(-1.639230\){\tiny \((\times 3)\)}; \(-1.646738\){\tiny \((\times 3)\)}; \(-2.000000\){\tiny \((\times 6)\)}; \(-2.400000\){\tiny \((\times 5)\)}; \(-3.200000\){\tiny \((\times 2)\)}; \(-3.313671\)
}\end{minipage}\\
\hline
\end{longtable}

\smallbreak\begin{longtable}{||l||}
\hline
\raise-0.1ex\hbox{
 \small {\bf S1165685}: \(V/g^2=-11.6568542495\), Quality \(-\log_{10}(|Q|^2)=31.35\) , \(U(1)\times U(1)\) }
\\
\hline
\hbox to6em{\(\phi\)}\quad\begin{minipage}{10.5cm}
{\vskip0.25ex
\small\tt
\(-0.764285_{1367s}\)  \(+0.764285_{2458s}\)  \(-0.764285_{1457c}\)  \(+0.764285_{1468c}\)  \(+0.764285_{2357c}\)  \(-0.764285_{2368c}\) 
}\end{minipage}\\
\hline
\hbox to6em{\((m^2/m_0^2)[\psi]\)}\quad\begin{minipage}{8.5cm}
{\vskip0.25ex
\small
\(+2.560660\){\tiny \((\times 8)\)}
}\end{minipage}\\
\hline
\hbox to6em{\((m^2/m_0^2)[\chi]\)}\quad\begin{minipage}{8.5cm}
{\vskip0.25ex
\small
\(+5.121320\){\tiny \((\times 8)\)}; \(+3.685461\){\tiny \((\times 8)\)}; \(+3.464406\){\tiny \((\times 8)\)}; \(+0.853614\){\tiny \((\times 8)\)}; \(+0.801529\){\tiny \((\times 8)\)}; \(+0.117840\){\tiny \((\times 8)\)}; \(+0.001772\){\tiny \((\times 8)\)}
}\end{minipage}\\
\hline
\hbox to6em{\((m/m_0)[\phi]\)}\quad\begin{minipage}{8.5cm}
{\vskip0.25ex
\small
\(+8.485281\){\tiny \((\times 2)\)}; \(+7.029437\); \(+6.363961\){\tiny \((\times 2)\)}; \(+5.651616\){\tiny \((\times 4)\)}; \(+4.801195\); \(+3.268128\){\tiny \((\times 4)\)}; \(+2.485281\); \(+0.000000\){\tiny \((\times 26)\)}; \(-0.514719\){\tiny \((\times 8)\)}; \(-0.782846\){\tiny \((\times 4)\)}; \(-1.091883\){\tiny \((\times 2)\)}; \(-1.286476\); \(-1.408975\){\tiny \((\times 4)\)}; \(-2.121320\){\tiny \((\times 4)\)}; \(-2.485281\); \(-3.514719\){\tiny \((\times 5)\)}
}\end{minipage}\\
\hline
\end{longtable}

\smallbreak\begin{longtable}{||l||}
\hline
\raise-0.1ex\hbox{
 \small {\bf S1176725}: \(V/g^2=-11.7672503174\), Quality \(-\log_{10}(|Q|^2)=33.22\) }
\\
\hline
\hbox to6em{\(\phi\)}\quad\begin{minipage}{10.5cm}
{\vskip0.25ex
\small\tt
\(+0.246799_{1238s}\)  \(+0.246812_{1246s}\)  \(+0.246799_{1257s}\)  \(+0.162463_{1267s}\)  \(+0.246799_{1345s}\)  \(-0.162463_{1346s}\)  \(+0.685377_{1367s}\)  \(-0.246799_{1478s}\)  \(+0.219283_{1568s}\)  \(+0.219283_{2347s}\)  \(-0.246799_{2356s}\)  \(-0.685377_{2458s}\)  \(-0.162463_{2578s}\)  \(-0.246799_{2678s}\)  \(+0.162463_{3458s}\)  \(-0.246799_{3468s}\)  \(-0.246812_{3578s}\)  \(+0.246799_{4567s}\)  \(-0.267710_{1236c}\)  \(+0.100175_{1248c}\)  \(-0.793675_{1278c}\)  \(-0.258255_{1348c}\)  \(-0.100175_{1378c}\)  \(-0.267710_{1467c}\)  \(+0.267710_{2358c}\)  \(-0.100175_{2456c}\)  \(+0.258255_{2567c}\)  \(+0.793675_{3456c}\)  \(+0.100175_{3567c}\)  \(+0.267710_{4578c}\) 
}\end{minipage}\\
\hline
\hbox to6em{\((m^2/m_0^2)[\psi]\)}\quad\begin{minipage}{8.5cm}
{\vskip0.25ex
\small
\(+2.831740\){\tiny \((\times 6)\)}; \(+2.154978\){\tiny \((\times 2)\)}
}\end{minipage}\\
\hline
\hbox to6em{\((m^2/m_0^2)[\chi]\)}\quad\begin{minipage}{8.5cm}
{\vskip0.25ex
\small
\(+5.663480\){\tiny \((\times 6)\)}; \(+4.841286\){\tiny \((\times 2)\)}; \(+4.340294\){\tiny \((\times 6)\)}; \(+4.309955\){\tiny \((\times 2)\)}; \(+3.906964\){\tiny \((\times 2)\)}; \(+3.001119\){\tiny \((\times 6)\)}; \(+0.960260\){\tiny \((\times 2)\)}; \(+0.910015\){\tiny \((\times 6)\)}; \(+0.862406\){\tiny \((\times 6)\)}; \(+0.326099\){\tiny \((\times 2)\)}; \(+0.232445\){\tiny \((\times 6)\)}; \(+0.122376\){\tiny \((\times 6)\)}; \(+0.109418\){\tiny \((\times 2)\)}; \(+0.001389\){\tiny \((\times 2)\)}
}\end{minipage}\\
\hline
\end{longtable}

\smallbreak\begin{longtable}{||l||}
\hline
\raise-0.1ex\hbox{
 \small {\bf S1195898}: \(V/g^2=-11.9589830843\), Quality \(-\log_{10}(|Q|^2)=21.12\) , dim(GG)=3}
\\
\hline
\hbox to6em{\(\phi\)}\quad\begin{minipage}{10.5cm}
{\vskip0.25ex
\small\tt
\(-0.231631_{1234s}\)  \(+0.119571_{1236s}\)  \(-0.246360_{1258s}\)  \(-0.246360_{1267s}\)  \(+0.246360_{1356s}\)  \(+0.246360_{1378s}\)  \(+0.724351_{1457s}\)  \(+0.246360_{1468s}\)  \(+0.119571_{1567s}\)  \(-0.119571_{2348s}\)  \(-0.246360_{2357s}\)  \(-0.724351_{2368s}\)  \(-0.246360_{2456s}\)  \(-0.246360_{2478s}\)  \(-0.246360_{3458s}\)  \(-0.246360_{3467s}\)  \(+0.119571_{4578s}\)  \(-0.231631_{5678s}\)  \(+0.240574_{1238c}\)  \(+0.132263_{1245c}\)  \(+0.240574_{1247c}\)  \(+0.222038_{1256c}\)  \(-0.240574_{1345c}\)  \(-0.132263_{1347c}\)  \(+0.703186_{1367c}\)  \(+0.240574_{1468c}\)  \(-0.240574_{1578c}\)  \(-0.240574_{2346c}\)  \(+0.240574_{2357c}\)  \(+0.703186_{2458c}\)  \(-0.132263_{2568c}\)  \(-0.240574_{2678c}\)  \(-0.222038_{3478c}\)  \(-0.240574_{3568c}\)  \(-0.132263_{3678c}\)  \(-0.240574_{4567c}\) 
}\end{minipage}\\
\hline
\hbox to6em{\((m^2/m_0^2)[\psi]\)}\quad\begin{minipage}{8.5cm}
{\vskip0.25ex
\small
\(+3.073009\){\tiny \((\times 3)\)}; \(+2.518511\){\tiny \((\times 5)\)}
}\end{minipage}\\
\hline
\hbox to6em{\((m^2/m_0^2)[\chi]\)}\quad\begin{minipage}{8.5cm}
{\vskip0.25ex
\small
\(+6.146019\){\tiny \((\times 3)\)}; \(+5.849903\); \(+5.037022\){\tiny \((\times 5)\)}; \(+4.805271\){\tiny \((\times 3)\)}; \(+3.998704\){\tiny \((\times 5)\)}; \(+3.396994\){\tiny \((\times 7)\)}; \(+1.331958\){\tiny \((\times 7)\)}; \(+0.415119\){\tiny \((\times 9)\)}; \(+0.391172\); \(+0.357233\){\tiny \((\times 3)\)}; \(+0.263746\){\tiny \((\times 7)\)}; \(+0.054804\){\tiny \((\times 5)\)}
}\end{minipage}\\
\hline
\end{longtable}

\smallbreak\begin{longtable}{||l||}
\hline
\raise-0.1ex\hbox{
 \small {\bf S1200000}: \(V/g^2=-12.0\), Quality \(-\log_{10}(|Q|^2)=26.75\) , \(U(1)\times U(1)\)  \(\mathcal{N}=1\) }
\\
\hline
\hbox to6em{\(\phi\)}\quad\begin{minipage}{10.5cm}
{\vskip0.25ex
\small\tt
\(+0.573108_{1367s}\)  \(-0.573108_{1458s}\)  \(-0.573108_{2367s}\)  \(-0.573108_{2458s}\)  \(-0.658479_{1278c}\)  \(-0.573108_{1367c}\)  \(+0.573108_{1458c}\)  \(-0.573108_{2367c}\)  \(-0.573108_{2458c}\)  \(+0.658479_{3456c}\) 
}\end{minipage}\\
\hline
\hbox to6em{\((m^2/m_0^2)[\psi]\)}\quad\begin{minipage}{8.5cm}
{\vskip0.25ex
\small
\(+4.000000\); \(+3.000000\){\tiny \((\times 2)\)}; \(+2.250000\){\tiny \((\times 4)\)}; \(+1.000000\)
}\end{minipage}\\
\hline
\hbox to6em{\((m^2/m_0^2)[\chi]\)}\quad\begin{minipage}{8.5cm}
{\vskip0.25ex
\small
\(+8.000000\); \(+6.000000\){\tiny \((\times 2)\)}; \(+4.746708\){\tiny \((\times 4)\)}; \(+4.500000\){\tiny \((\times 4)\)}; \(+3.936492\){\tiny \((\times 4)\)}; \(+3.732051\); \(+2.165569\){\tiny \((\times 4)\)}; \(+2.000000\){\tiny \((\times 5)\)}; \(+1.500000\){\tiny \((\times 2)\)}; \(+1.125000\){\tiny \((\times 4)\)}; \(+0.584431\){\tiny \((\times 4)\)}; \(+0.500000\){\tiny \((\times 4)\)}; \(+0.267949\); \(+0.125000\){\tiny \((\times 4)\)}; \(+0.063508\){\tiny \((\times 4)\)}; \(+0.003292\){\tiny \((\times 4)\)}; \(+0.000000\){\tiny \((\times 4)\)}
}\end{minipage}\\
\hline
\hbox to6em{\((m/m_0)[\phi]\)}\quad\begin{minipage}{8.5cm}
{\vskip0.25ex
\small
\(+8.678867\){\tiny \((\times 4)\)}; \(+8.196152\); \(+4.412278\){\tiny \((\times 4)\)}; \(+3.067100\){\tiny \((\times 4)\)}; \(+2.732051\); \(+0.000000\){\tiny \((\times 30)\)}; \(-0.732051\); \(-1.250000\){\tiny \((\times 4)\)}; \(-1.516589\){\tiny \((\times 4)\)}; \(-1.912278\){\tiny \((\times 4)\)}; \(-2.000000\){\tiny \((\times 4)\)}; \(-2.196152\); \(-2.229377\){\tiny \((\times 4)\)}; \(-2.250000\){\tiny \((\times 4)\)}
}\end{minipage}\\
\hline
\end{longtable}

\smallbreak\begin{longtable}{||l||}
\hline
\raise-0.1ex\hbox{
 \small {\bf S1212986}: \(V/g^2=-12.1298657377\), Quality \(-\log_{10}(|Q|^2)=27.73\) , \(U(1)\) }
\\
\hline
\hbox to6em{\(\phi\)}\quad\begin{minipage}{10.5cm}
{\vskip0.25ex
\small\tt
\(+0.235254_{1234s}\)  \(-0.254267_{1258s}\)  \(+0.254267_{1267s}\)  \(+0.306918_{1357s}\)  \(+0.306918_{1368s}\)  \(-0.714110_{1456s}\)  \(-0.205576_{1478s}\)  \(-0.205576_{2356s}\)  \(-0.714110_{2378s}\)  \(+0.306918_{2457s}\)  \(+0.306918_{2468s}\)  \(+0.254267_{3458s}\)  \(-0.254267_{3467s}\)  \(+0.235254_{5678s}\)  \(-0.271494_{1256c}\)  \(+0.271494_{1278c}\)  \(+0.816935_{1358c}\)  \(+0.273947_{1367c}\)  \(+0.273947_{2458c}\)  \(+0.816935_{2467c}\)  \(-0.271494_{3456c}\)  \(+0.271494_{3478c}\) 
}\end{minipage}\\
\hline
\hbox to6em{\((m^2/m_0^2)[\psi]\)}\quad\begin{minipage}{8.5cm}
{\vskip0.25ex
\small
\(+3.535597\){\tiny \((\times 2)\)}; \(+2.920664\){\tiny \((\times 2)\)}; \(+2.398064\){\tiny \((\times 2)\)}; \(+2.276432\){\tiny \((\times 2)\)}
}\end{minipage}\\
\hline
\end{longtable}

\smallbreak\begin{longtable}{||l||}
\hline
\raise-0.1ex\hbox{
 \small {\bf S1271622}: \(V/g^2=-12.716224888\), Quality \(-\log_{10}(|Q|^2)=29.31\) , dim(GG)=3}
\\
\hline
\hbox to6em{\(\phi\)}\quad\begin{minipage}{10.5cm}
{\vskip0.25ex
\small\tt
\(-0.417600_{1248s}\)  \(-0.417600_{1257s}\)  \(-0.417600_{1456s}\)  \(+0.417600_{1678s}\)  \(+0.417600_{2345s}\)  \(-0.417600_{2378s}\)  \(+0.417600_{3468s}\)  \(+0.417600_{3567s}\)  \(-0.049912_{1234c}\)  \(-0.794284_{1256c}\)  \(-0.794284_{1367c}\)  \(-0.049912_{1457c}\)  \(-0.049912_{2368c}\)  \(-0.794284_{2458c}\)  \(+0.794284_{3478c}\)  \(+0.049912_{5678c}\) 
}\end{minipage}\\
\hline
\hbox to6em{\((m^2/m_0^2)[\psi]\)}\quad\begin{minipage}{8.5cm}
{\vskip0.25ex
\small
\(+3.594847\){\tiny \((\times 4)\)}; \(+2.012973\){\tiny \((\times 4)\)}
}\end{minipage}\\
\hline
\hbox to6em{\((m^2/m_0^2)[\chi]\)}\quad\begin{minipage}{8.5cm}
{\vskip0.25ex
\small
\(+7.189693\){\tiny \((\times 4)\)}; \(+5.142313\){\tiny \((\times 4)\)}; \(+4.631511\){\tiny \((\times 4)\)}; \(+4.025947\){\tiny \((\times 4)\)}; \(+3.596011\){\tiny \((\times 4)\)}; \(+2.797159\){\tiny \((\times 4)\)}; \(+1.676916\){\tiny \((\times 4)\)}; \(+1.471722\){\tiny \((\times 8)\)}; \(+0.350381\){\tiny \((\times 4)\)}; \(+0.064058\){\tiny \((\times 4)\)}; \(+0.019667\){\tiny \((\times 8)\)}; \(+0.013615\){\tiny \((\times 4)\)}
}\end{minipage}\\
\hline
\hbox to6em{\((m/m_0)[\phi]\)}\quad\begin{minipage}{8.5cm}
{\vskip0.25ex
\small
\(+9.373526\); \(+8.326452\){\tiny \((\times 2)\)}; \(+8.107084\); \(+7.297178\){\tiny \((\times 3)\)}; \(+7.247748\); \(+6.291101\){\tiny \((\times 6)\)}; \(+2.383617\); \(+1.288334\){\tiny \((\times 3)\)}; \(+0.000000\){\tiny \((\times 25)\)}; \(-1.055457\){\tiny \((\times 6)\)}; \(-1.172410\){\tiny \((\times 3)\)}; \(-1.362824\){\tiny \((\times 6)\)}; \(-1.483932\); \(-1.714334\); \(-1.864898\){\tiny \((\times 3)\)}; \(-1.931500\); \(-2.625043\){\tiny \((\times 5)\)}; \(-3.353494\)
}\end{minipage}\\
\hline
\end{longtable}

\smallbreak\begin{longtable}{||l||}
\hline
\raise-0.1ex\hbox{
 \small {\bf S1301601}: \(V/g^2=-13.0160180374\), Quality \(-\log_{10}(|Q|^2)=21.08\) , \(U(1)\times U(1)\) }
\\
\hline
\hbox to6em{\(\phi\)}\quad\begin{minipage}{10.5cm}
{\vskip0.25ex
\small\tt
\(-0.323726_{1236s}\)  \(-0.323726_{1245s}\)  \(-0.497580_{1246s}\)  \(+0.231176_{1278s}\)  \(+0.266403_{1348s}\)  \(+0.323726_{1357s}\)  \(-0.497580_{1457s}\)  \(-0.323726_{1467s}\)  \(-0.231176_{1568s}\)  \(-0.231176_{2347s}\)  \(-0.323726_{2358s}\)  \(+0.497580_{2368s}\)  \(+0.323726_{2468s}\)  \(+0.266403_{2567s}\)  \(+0.231176_{3456s}\)  \(+0.497580_{3578s}\)  \(-0.323726_{3678s}\)  \(-0.323726_{4578s}\)  \(-0.958766_{1367c}\)  \(-0.958766_{2458c}\) 
}\end{minipage}\\
\hline
\hbox to6em{\((m^2/m_0^2)[\psi]\)}\quad\begin{minipage}{8.5cm}
{\vskip0.25ex
\small
\(+3.005710\){\tiny \((\times 8)\)}
}\end{minipage}\\
\hline
\hbox to6em{\((m^2/m_0^2)[\chi]\)}\quad\begin{minipage}{8.5cm}
{\vskip0.25ex
\small
\(+6.011420\){\tiny \((\times 8)\)}; \(+5.321843\){\tiny \((\times 8)\)}; \(+4.126568\){\tiny \((\times 8)\)}; \(+0.855344\){\tiny \((\times 16)\)}; \(+0.784661\){\tiny \((\times 8)\)}; \(+0.096209\){\tiny \((\times 8)\)}
}\end{minipage}\\
\hline
\hbox to6em{\((m/m_0)[\phi]\)}\quad\begin{minipage}{8.5cm}
{\vskip0.25ex
\small
\(+14.518509\); \(+12.953583\){\tiny \((\times 2)\)}; \(+11.056122\); \(+10.008967\); \(+7.458673\); \(+4.469179\){\tiny \((\times 4)\)}; \(+3.198206\){\tiny \((\times 4)\)}; \(+3.025692\){\tiny \((\times 4)\)}; \(+2.913149\){\tiny \((\times 4)\)}; \(+0.000000\){\tiny \((\times 26)\)}; \(-0.214642\); \(-0.514678\); \(-0.647720\); \(-1.083233\){\tiny \((\times 4)\)}; \(-1.478523\){\tiny \((\times 4)\)}; \(-1.716653\); \(-1.909115\){\tiny \((\times 4)\)}; \(-2.590044\){\tiny \((\times 4)\)}; \(-3.590693\); \(-3.806792\)
}\end{minipage}\\
\hline
\end{longtable}

\smallbreak\begin{longtable}{||l||}
\hline
\raise-0.1ex\hbox{
 \small {\bf S1362365}: \(V/g^2=-13.6236525917\), Quality \(-\log_{10}(|Q|^2)=18.70\) , \(U(1)\) }
\\
\hline
\hbox to6em{\(\phi\)}\quad\begin{minipage}{10.5cm}
{\vskip0.25ex
\small\tt
\(+0.351118_{1234s}\)  \(+0.501138_{1256s}\)  \(-0.153800_{1278s}\)  \(+0.018756_{1358s}\)  \(+0.493282_{1367s}\)  \(-0.493282_{1457s}\)  \(+0.018756_{1468s}\)  \(-0.018756_{2357s}\)  \(+0.493282_{2368s}\)  \(-0.493282_{2458s}\)  \(-0.018756_{2467s}\)  \(-0.153800_{3456s}\)  \(+0.501138_{3478s}\)  \(+0.351118_{5678s}\)  \(+0.130896_{1234c}\)  \(+0.553052_{1256c}\)  \(+0.386221_{1278c}\)  \(-0.172818_{1358c}\)  \(+0.578894_{1367c}\)  \(-0.578894_{1457c}\)  \(-0.172818_{1468c}\)  \(-0.172818_{2357c}\)  \(-0.578894_{2368c}\)  \(+0.578894_{2458c}\)  \(-0.172818_{2467c}\)  \(-0.386221_{3456c}\)  \(-0.553052_{3478c}\)  \(-0.130896_{5678c}\) 
}\end{minipage}\\
\hline
\hbox to6em{\((m^2/m_0^2)[\psi]\)}\quad\begin{minipage}{8.5cm}
{\vskip0.25ex
\small
\(+3.424801\); \(+3.391334\); \(+3.377415\); \(+3.080922\){\tiny \((\times 2)\)}; \(+2.306966\){\tiny \((\times 2)\)}; \(+1.430973\)
}\end{minipage}\\
\hline
\end{longtable}

\smallbreak\begin{longtable}{||l||}
\hline
\raise-0.1ex\hbox{
 \small {\bf S1363782}: \(V/g^2=-13.6378280148\), Quality \(-\log_{10}(|Q|^2)=32.39\) }
\\
\hline
\hbox to6em{\(\phi\)}\quad\begin{minipage}{10.5cm}
{\vskip0.25ex
\small\tt
\(+0.049498_{1234s}\)  \(-0.075759_{1237s}\)  \(-0.005984_{1245s}\)  \(+0.054500_{1246s}\)  \(+0.072193_{1258s}\)  \(-0.000963_{1267s}\)  \(-0.001279_{1268s}\)  \(+0.144635_{1345s}\)  \(+0.005984_{1346s}\)  \(+0.000963_{1357s}\)  \(-0.001279_{1358s}\)  \(+0.060130_{1368s}\)  \(-0.049498_{1456s}\)  \(-0.028514_{1478s}\)  \(+0.920210_{1567s}\)  \(-0.920210_{2348s}\)  \(-0.028514_{2356s}\)  \(-0.049498_{2378s}\)  \(+0.060130_{2457s}\)  \(+0.001279_{2467s}\)  \(+0.000963_{2468s}\)  \(+0.005984_{2578s}\)  \(-0.144635_{2678s}\)  \(+0.001279_{3457s}\)  \(-0.000963_{3458s}\)  \(+0.072193_{3467s}\)  \(-0.054500_{3578s}\)  \(-0.005984_{3678s}\)  \(+0.075759_{4568s}\)  \(+0.049498_{5678s}\)  \(-0.011261_{1234c}\)  \(+0.228668_{1238c}\)  \(+0.049153_{1245c}\)  \(+0.108395_{1246c}\)  \(-0.212717_{1257c}\)  \(+0.003275_{1267c}\)  \(+0.000745_{1268c}\)  \(-0.378219_{1345c}\)  \(-0.049153_{1346c}\)  \(+0.003275_{1357c}\)  \(-0.000745_{1358c}\)  \(+0.920224_{1367c}\)  \(+0.011261_{1456c}\)  \(+0.234756_{1478c}\)  \(+0.204611_{1568c}\)  \(-0.204611_{2347c}\)  \(-0.234756_{2356c}\)  \(-0.011261_{2378c}\)  \(+0.920224_{2458c}\)  \(-0.000745_{2467c}\)  \(-0.003275_{2468c}\)  \(+0.049153_{2578c}\)  \(-0.378219_{2678c}\)  \(+0.000745_{3457c}\)  \(-0.003275_{3458c}\)  \(-0.212717_{3468c}\)  \(+0.108395_{3578c}\)  \(-0.049153_{3678c}\)  \(-0.228668_{4567c}\)  \(+0.011261_{5678c}\) 
}\end{minipage}\\
\hline
\hbox to6em{\((m^2/m_0^2)[\psi]\)}\quad\begin{minipage}{8.5cm}
{\vskip0.25ex
\small
\(+3.523713\); \(+3.428003\); \(+3.399770\); \(+3.061453\); \(+2.975587\); \(+2.478208\); \(+2.118747\); \(+1.500004\)
}\end{minipage}\\
\hline
\end{longtable}

\smallbreak\begin{longtable}{||l||}
\hline
\raise-0.1ex\hbox{
 \small {\bf S1366864}: \(V/g^2=-13.6686490616\), Quality \(-\log_{10}(|Q|^2)=32.28\) }
\\
\hline
\hbox to6em{\(\phi\)}\quad\begin{minipage}{10.5cm}
{\vskip0.25ex
\small\tt
\(+0.096839_{1234s}\)  \(+0.082235_{1258s}\)  \(+0.053280_{1267s}\)  \(-0.930679_{1356s}\)  \(-0.082235_{1378s}\)  \(+0.096839_{1457s}\)  \(+0.070464_{1468s}\)  \(-0.070464_{2357s}\)  \(-0.096839_{2368s}\)  \(+0.082235_{2456s}\)  \(+0.930679_{2478s}\)  \(+0.053280_{3458s}\)  \(+0.082235_{3467s}\)  \(+0.096839_{5678s}\)  \(-0.307147_{1238c}\)  \(+0.213591_{1256c}\)  \(-0.211547_{1278c}\)  \(+0.211547_{1358c}\)  \(-0.929931_{1367c}\)  \(-0.307147_{1578c}\)  \(-0.307147_{2346c}\)  \(-0.929931_{2458c}\)  \(+0.211547_{2467c}\)  \(+0.211547_{3456c}\)  \(-0.213591_{3478c}\)  \(+0.307147_{4567c}\) 
}\end{minipage}\\
\hline
\hbox to6em{\((m^2/m_0^2)[\psi]\)}\quad\begin{minipage}{8.5cm}
{\vskip0.25ex
\small
\(+3.456697\){\tiny \((\times 2)\)}; \(+3.264549\){\tiny \((\times 2)\)}; \(+2.715801\){\tiny \((\times 2)\)}; \(+1.773512\){\tiny \((\times 2)\)}
}\end{minipage}\\
\hline
\end{longtable}

\smallbreak\begin{longtable}{||l||}
\hline
\raise-0.1ex\hbox{
 \small {\bf S1367611}: \(V/g^2=-13.6761142184\), Quality \(-\log_{10}(|Q|^2)=21.31\) }
\\
\hline
\hbox to6em{\(\phi\)}\quad\begin{minipage}{10.5cm}
{\vskip0.25ex
\small\tt
\(+0.102432_{1238s}\)  \(+0.056950_{1247s}\)  \(-0.052291_{1248s}\)  \(+0.082782_{1256s}\)  \(+0.082782_{1345s}\)  \(+0.921226_{1367s}\)  \(+0.052291_{1368s}\)  \(-0.102432_{1468s}\)  \(-0.086344_{1578s}\)  \(+0.086344_{2346s}\)  \(+0.102432_{2357s}\)  \(+0.052291_{2457s}\)  \(-0.921226_{2458s}\)  \(-0.082782_{2678s}\)  \(+0.082782_{3478s}\)  \(+0.052291_{3567s}\)  \(+0.056950_{3568s}\)  \(+0.102432_{4567s}\)  \(+0.327140_{1235c}\)  \(+0.225595_{1245c}\)  \(-0.210554_{1267c}\)  \(-0.053658_{1268c}\)  \(-0.916874_{1347c}\)  \(+0.053658_{1348c}\)  \(+0.225595_{1356c}\)  \(+0.327140_{1456c}\)  \(-0.327140_{2378c}\)  \(+0.225595_{2478c}\)  \(-0.053658_{2567c}\)  \(-0.916874_{2568c}\)  \(-0.053658_{3457c}\)  \(+0.210554_{3458c}\)  \(-0.225595_{3678c}\)  \(+0.327140_{4678c}\) 
}\end{minipage}\\
\hline
\hbox to6em{\((m^2/m_0^2)[\psi]\)}\quad\begin{minipage}{8.5cm}
{\vskip0.25ex
\small
\(+3.619733\); \(+3.338988\){\tiny \((\times 2)\)}; \(+3.219778\); \(+2.973868\); \(+2.612441\); \(+1.797021\){\tiny \((\times 2)\)}
}\end{minipage}\\
\hline
\end{longtable}

\smallbreak\begin{longtable}{||l||}
\hline
\raise-0.1ex\hbox{
 \small {\bf S1379439}: \(V/g^2=-13.7943987074\), Quality \(-\log_{10}(|Q|^2)=21.14\) }
\\
\hline
\hbox to6em{\(\phi\)}\quad\begin{minipage}{10.5cm}
{\vskip0.25ex
\small\tt
\(-0.234229_{1236s}\)  \(+0.218874_{1248s}\)  \(-0.180314_{1257s}\)  \(+0.133102_{1268s}\)  \(-0.234229_{1347s}\)  \(-0.187620_{1358s}\)  \(-0.180314_{1456s}\)  \(+0.133102_{1478s}\)  \(-0.620183_{1678s}\)  \(-0.620183_{2345s}\)  \(+0.133102_{2356s}\)  \(-0.180314_{2378s}\)  \(+0.187620_{2467s}\)  \(+0.234229_{2568s}\)  \(-0.133102_{3457s}\)  \(+0.180314_{3468s}\)  \(-0.218874_{3567s}\)  \(-0.234229_{4578s}\)  \(-0.055695_{1245c}\)  \(-0.008952_{1256c}\)  \(+0.032098_{1278c}\)  \(+0.825302_{1457c}\)  \(+0.912083_{1468c}\)  \(+0.055695_{1567c}\)  \(+0.055695_{2348c}\)  \(+0.912083_{2357c}\)  \(+0.825302_{2368c}\)  \(-0.032098_{3456c}\)  \(+0.008952_{3478c}\)  \(+0.055695_{3678c}\) 
}\end{minipage}\\
\hline
\hbox to6em{\((m^2/m_0^2)[\psi]\)}\quad\begin{minipage}{8.5cm}
{\vskip0.25ex
\small
\(+4.367804\); \(+3.625737\){\tiny \((\times 2)\)}; \(+3.157296\); \(+2.492601\){\tiny \((\times 2)\)}; \(+2.394818\); \(+1.523914\)
}\end{minipage}\\
\hline
\end{longtable}

\smallbreak\begin{longtable}{||l||}
\hline
\raise-0.1ex\hbox{
 \small {\bf S1384135}: \(V/g^2=-13.841358302\), Quality \(-\log_{10}(|Q|^2)=20.96\) , \(U(1)\) }
\\
\hline
\hbox to6em{\(\phi\)}\quad\begin{minipage}{10.5cm}
{\vskip0.25ex
\small\tt
\(+0.161146_{1234s}\)  \(+0.185149_{1257s}\)  \(-0.006072_{1268s}\)  \(-0.185149_{1358s}\)  \(+0.752924_{1367s}\)  \(+0.351922_{1456s}\)  \(+0.161146_{1478s}\)  \(+0.161146_{2356s}\)  \(+0.351922_{2378s}\)  \(-0.752924_{2458s}\)  \(+0.185149_{2467s}\)  \(+0.006072_{3457s}\)  \(-0.185149_{3468s}\)  \(+0.161146_{5678s}\)  \(-0.155065_{1234c}\)  \(+0.318508_{1258c}\)  \(+0.165129_{1267c}\)  \(+0.893256_{1357c}\)  \(-0.165129_{1368c}\)  \(+0.581767_{1456c}\)  \(-0.155065_{1478c}\)  \(+0.155065_{2356c}\)  \(-0.581767_{2378c}\)  \(+0.165129_{2457c}\)  \(-0.893256_{2468c}\)  \(-0.165129_{3458c}\)  \(-0.318508_{3467c}\)  \(+0.155065_{5678c}\) 
}\end{minipage}\\
\hline
\hbox to6em{\((m^2/m_0^2)[\psi]\)}\quad\begin{minipage}{8.5cm}
{\vskip0.25ex
\small
\(+3.425450\){\tiny \((\times 2)\)}; \(+3.386097\); \(+3.184549\); \(+2.933695\); \(+2.887794\){\tiny \((\times 2)\)}; \(+1.006714\)
}\end{minipage}\\
\hline
\end{longtable}

\smallbreak\begin{longtable}{||l||}
\hline
\raise-0.1ex\hbox{
 \small {\bf S1400000}: \(V/g^2=-14.0\), Quality \(-\log_{10}(|Q|^2)=106.43\) , dim(GG)=6}
\\
\hline
\hbox to6em{\(\phi\)}\quad\begin{minipage}{10.5cm}
{\vskip0.25ex
\small\tt
\(+1.020804_{1235s}\)  \(-1.020804_{4678s}\)  \(+1.020804_{1234c}\)  \(-1.020804_{5678c}\) 
}\end{minipage}\\
\hline
\hbox to6em{\((m^2/m_0^2)[\psi]\)}\quad\begin{minipage}{8.5cm}
{\vskip0.25ex
\small
\(+3.857143\){\tiny \((\times 2)\)}; \(+2.142857\){\tiny \((\times 6)\)}
}\end{minipage}\\
\hline
\hbox to6em{\((m^2/m_0^2)[\chi]\)}\quad\begin{minipage}{8.5cm}
{\vskip0.25ex
\small
\(+7.714286\){\tiny \((\times 2)\)}; \(+4.285714\){\tiny \((\times 12)\)}; \(+2.244029\){\tiny \((\times 18)\)}; \(+0.327399\){\tiny \((\times 18)\)}; \(+0.000000\){\tiny \((\times 6)\)}
}\end{minipage}\\
\hline
\hbox to6em{\((m/m_0)[\phi]\)}\quad\begin{minipage}{8.5cm}
{\vskip0.25ex
\small
\(+8.571429\); \(+5.598110\){\tiny \((\times 9)\)}; \(+2.571429\){\tiny \((\times 9)\)}; \(+0.000000\){\tiny \((\times 22)\)}; \(-1.312396\){\tiny \((\times 9)\)}; \(-1.714286\){\tiny \((\times 20)\)}
}\end{minipage}\\
\hline
\end{longtable}%

\subsection{Version 1}

\paragraph{Preamble}
This subsection discusses special properties of solutions known at the
release of version~1 of this part of the collection. Subsequent
incremental updates will be added in subsections titled ``Version 2''
etc. Version numbers of the four parts of the encyclopedic reference
evolve independently from one another. As discussed in the explanation
at the beginning of part~1 of the series, the discussion deliberately
avoids global statements along the lines of ``this is the only known
solution with property X'' that may be invalidated by subsequent
updates of this, or any other, part of the collection.

{\bf S0600000} is the solution at the origin with unbroken
$\mathcal{N}=8$ supersymmetry. The existence of this solution follows
directly from the expressions given in the article by de Wit and
Nicolai that first constructed $SO(8)$-gauged
supergravity~\xcite{de Wit:1982ig}. It first is discussed in detail
in~\xcite{Warner:1983du}. It is by no means true that all gauged
maximal supergravities have such a trivial critical point. For
example, $\mathcal{N}=8$ supergravity also permits the gauging of
other compact forms of $SO(8)$ and contractions thereof
(cf.~\xcite{Hull:1984qz}), and the $SO(7,1)$-gauged model does not have a
$\phi=0$ stationary point~\xcite{Hull:1984ea}.

{\bf S0668740} and {\bf S0698771} are unstable critical points
with remarkable properties. They can be obtained by picking a single
$SO(8)$~spinor (respectively co-spinor), forming a symmetric traceless
matrix from it and the identity, and setting the co-spinorial
(respectively spinorial) part of the 70-vector to zero. These
solutions have matching mass spectra, but different cosmological
constants: exchanging ${\bf 35}_s\leftrightarrow {\bf 35}_c$ is not a
symmetry of $SO(8)$-gauged $\mathcal{N}=8$ supergravity. Lifting these
to 11-dimensional supergravity, these solutions correspond to
compactifications found by Francois Englert in~\xcite{Englert:1982vs}.
Details of these solutions are discussed
in~\xcite{deWit:1984nz,deWit:1984va,Englert:1983jc,Biran:1984jr,Warner:1983vz}.

{\bf S0719157} is a stable $\mathcal{N}=1$ supersymmetric vacuum
with residual gauge symmetry $G_2$. This has been discussed 
in~\xcite{Warner:1983du,deWit:1984nz}. Renormalization group flows that
connect this solution with the $SO(8)$-symmetric $\mathcal{N}=8$
solution {\bf S0600000} and the $SU(3)\times U(1)$-symmetric 
$\mathcal{N}=2$ solution {\bf S779422} have been discussed in the
context of $M2$-brane field theory in~\xcite{Bobev:2009ms}.

{\bf S0779422} is a stable $\mathcal{N}=2$ supersymmetric vacuum
with residual gauge symmetry $SU(3)\times U(1)$. This solution has
first been described in~\xcite{Warner:1983vz} and was the object of
numerous investigations, in part due to speculations of its potential
phenomenological significance~\xcite{Nicolai:1985hs} (see also
\xcite{Nicolai:2010zza}), as well as its relevance to M2 brane physics
and ABJM/BLG theory (cf.~\xcite{Benna:2008zy,Klebanov:2008vq,Klebanov:2009kp,Corrado:2001nv,Bobev:2009ms,Ahn:2000aq,Ahn:2010zi}).

{\bf S0800000} is an unstable critical point with {\em integer}
cosmological constant $V/g^2=-8$ and residual gauge group $SU(4)$. It
is noteworthy that only scalars in the ${\bf 35}_c$ part of $\phi$ are
non-zero here. This solution was first found in~\xcite{Warner:1983vz}
and is also studied in~\xcite{Pope:1984bd}. The scalar mass spectrum as
well as the significance of the instability of this and other critical
points to AdS/CMT are discussed in~\xcite{Bobev:2010ib}.

{\bf S0847213} is a new unstable critical point with
five-dimensional residual gauge group. As this group cannot be
$U(1)^5$ (as $SO(8)$ is of rank~4), it must be $SO(3)\times U(1)\times
U(1)$. This is also suggested by scalar mass degeneracies. Inverse
symbolic computation using the PSLQ
algorithm~\xcite{PSLQ} on precision
data strongly suggests that the algebraic expression for the
cosmological constant is~$-\sqrt{36+16\sqrt{5}}$, and analytic
expressions for the location of this solution are 
\(\phi_{1236s}=\phi_{1678s}=\phi_{2345s}=
 \phi_{4578s}=\phi_{1268c}=\phi_{1367c}=\phi_{2458c}=\phi_{3457c}=
 -\frac{1}{4}\log\left(1+\frac{4}{5}\sqrt{5}-2\sqrt{\frac{2}{5}\left(2+\sqrt{5}\right)}\right)
\)
and \(\phi_{1456s}=\phi_{2378s}=\frac{1}{4}\log 5\).

{\bf S0869596} is a new unstable critical point with
four-dimensional residual gauge group, likely (considering mass
degeneracies) $SO(3)\times U(1)$. A precision calculation strongly
suggests that the algebraic expression for the cosmological constant
is~$-\frac{4}{5}\sqrt{54 + 14\sqrt{21}}$. Inverse symbolic
computation suggests these analytic expressions for the location of
the solution:
%
% +0.450223 1237s -0.450223 1246s -0.450223 1258s -0.450223 3467s
% +0.450223 3578s -0.450223 4568s -0.167520 1345c +0.167520 1367c
% +0.502560 1478c +0.167520 1568c -0.167520 2347c -0.502560 2356c
% +0.167520 2458c -0.167520 2678c
%
% A: 0.450223
\(\phi_{1237s}=
-\phi_{1246s}=
-\phi_{1258s}=
-\phi_{3467s}=
\phi_{3578s}=
-\phi_{4568s}=
\frac{1}{4}\log\left(\frac{1}{3}\left(9+2\sqrt{21}\right)\right)
\),
% B: 0.1675198695787
\(
-\phi_{1345c}=
\phi_{1367c}=
\phi_{1568c}=
-\phi_{2347c}=
\phi_{2458c}=
-\phi_{2678c}=
\frac{1}{4}\log\left(\frac{1}{5}\left(-3+2\sqrt{21}+2\sqrt{17-3\sqrt{21}}\right)\right)
\),
% C: 0.5025596
\(
\phi_{1478c}=
-\phi_{2356c}=
\frac{1}{4}\log\left(\frac{1}{125}\left(-2907+738\sqrt{21}+2\sqrt{4968137-1072683\sqrt{21}}\right)\right)
\)
Mass spectra suggest a strong similarity between this solution and
{\bf S0983994}.

{\bf S880733} has residual $SO(3)\times SO(3)$ gauge symmetry.
This unstable solution was first described in some detail in
\xcite{Fischbacher:2010ec}; the cosmological constant 
is~$-6\sqrt{1+\frac{2}{3}\sqrt{3}}$.

{\bf S0983994} is another novel (unstable) solution with
4-dimensional residual gauge symmetry which again is likely to be
$SO(3)\times U(1)$. Mass spectra suggest a strong similarity between
this solution and {\bf S0869596}. Curiously, a deep scan over the
scalar potentials using the method described in~\xcite{Fischbacher:2009cj}
with more than 4000~successful trial runs never managed to produce
this critical point, despite the search being most intensive around
$-V/g\sim 10$. This can be seen as a strong indication that the search
strategy to use standard numerical minimizers on the Frobenius norm of
the misalignment tensor~$|Q|^2$ (cf.~\xcite{Fischbacher:2009cj}) seems to
systematically miss some solutions. Eventually, this critical point
was obtained using a modification of the search strategy described in
the discussion of~{\bf S1400000}. The cosmological constant very
likely is $-5\cdot15^{1/4}$.

{\bf S0998708} is an unstable critical point with residual
$U(1)$ symmetry that was first found in \xcite{Fischbacher:2009cj}, where it
is given as solution \#7.

{\bf S?} A solution with cosmological constant $V/g^2=-10$ might
exist, considering that the scalar potential is now known to have
critical points with integer cosmological constants $V/g^2\in\{-6, -8,
  -12, -14, -16, -18\}$. As the discussion of {\bf S0983994} and {\bf
  S1400000} shows, the search method employed to find the first~$100+$
solutions that became known does seem to systematically miss some
critical points.

{\bf S1006758} is a novel unstable solution with residual gauge
symmetry $U(1)$.

{\bf S1039624} is a novel unstable solution with residual gauge
symmetry $U(1)$.

{\bf S1043471} is an unstable solution that breaks all gauge
symmetry. This was first described in~\xcite{Fischbacher:2009cj} as
solution~\#8. Numerical information about properties of this critical
point is given to 150-digit accuracy in~\xcite{Fischbacher:2010ec}. 

{\bf S1046017} is a novel unstable solution without any residual
gauge symmetry.

{\bf S1067475} is an unstable solution with residual $U(1)\times
U(1)$ gauge symmetry. This was first described in~\xcite{Fischbacher:2009cj}
as solution~\#9. Numerical information about properties of this
critical point is given to 150-digit accuracy
in~\xcite{Fischbacher:2010ec}, 
where an analytic expression for the cosmological constant is also 
given that was obtained by inverse symbolic calculation. It is:
\[
\begin{array}{lcl}
-V/g^2 &=& \left(\frac{QW^2+RW+S}{QW}\right)^{1/4}\\
\mbox{where}&&\\
Q&=&6561\\
R&=&28482192\\
S&=&122545537024\\
W&=&((128692865796145152+20596696547328\sqrt{2343}\,i)/1594323)^{1/3}
\end{array}
\]

{\bf S1068971} is a novel unstable solution with residual $U(1)\times
U(1)$ gauge symmetry.

{\bf S1075828} is a novel unstable solution with residual
four-dimensional gauge symmetry; considering mass degeneracies, this
is presumably $SO(3)\times U(1)$. 

{\bf S1165685} is an unstable solution with residual $U(1)\times
U(1)$ gauge symmetry. This was first described
in~\xcite{Fischbacher:2009cj} as solution~\#10.

{\bf S1176725} is a novel unstable solution without any residual
gauge symmetry.

{\bf S1195898} is a novel unstable solution with
three-dimensional residual gauge symmetry. Mass degeneracies strongly
suggest this to be $SO(3)$. This solution seems to share a number of
properties with another newly discovered solution, {\bf S2503105}.

{\bf S1200000} is the $\mathcal{N}=1$ supersymmetric vacuum with
residual gauge symmetry $U(1)\times U(1)$ that saturates the
Breitenlohner-Freedman bound~\xcite{Breitenlohner:1982jf,Breitenlohner:1982bm}. This was first discovered
in~\xcite{Fischbacher:2009cj} and discussed in detail 
in~\xcite{Fischbacher:2010ec}.

{\bf S1212986} is a novel unstable solution with residual gauge
symmetry $U(1)$.

{\bf S1271622} is a novel unstable solution with residual gauge
symmetry (likely) $SO(3)$.

{\bf S1301601} is a novel unstable solution with residual gauge
symmetry $U(1)\times U(1)$ and a remarkable gravitino mass spectrum.

{\bf S1362365} is an unstable solution with residual gauge symmetry
$U(1)$ that was first described in~\xcite{Fischbacher:2009cj} as
solution~\#12. This also is one of the solutions that only could be
obtained with a modified optimization strategy (minimizing not $|Q|^2$
but instead $|\nabla P|^2$ via double algorithmic differentiation).

{\bf S1363782} is a novel unstable solution without any residual gauge
symmetry.

{\bf S1366864} is a novel unstable solution without any residual gauge
symmetry.

{\bf S1367611} is an unstable solution without any residual gauge
symmetry that was first described in~\xcite{Fischbacher:2009cj} as
solution~\#13.

{\bf S1379439} is a novel unstable solution without any residual gauge
symmetry.

{\bf S1384135} is a fairly remarkable novel solution with residual
gauge symmetry $U(1)$, that {\em just} violates both the conditions
for supersymmetry and stability. 

{\bf S1400000} is the $SO(3)\times SO(3)$ {\emph stable} critical
point that was first discovered and described in~\xcite{Warner:1983du},
and indeed the first critical point with broken symmetry to be
described for four-dimensional $\mathcal{N}=8$
supergravity. The cosmological constant is exactly~$V/g^2=-14$, and
the location is given
by~\(\phi_{1235s}=-\phi_{4678s}=\phi_{1234c}=-\phi_{5678c}=\frac{1}{2}\sqrt{2}\,{\rm atanh}\,\left(\frac{2\sqrt{5}}{5}\right)\).

Remarkably, all the other non-supersymmetric critical
points in maximal supergravities that have ever been found in the
pre-2000 era violate the Breitenlohner-Freedman
bound~\xcite{Breitenlohner:1982jf,Breitenlohner:1982bm} and are hence
perturbatively unstable. This gave rise to the belief that stability
and supersymmetry were closely linked. It was only in~2010 that the
full scalar mass spectrum of this solution was
computed~\xcite{Fischbacher:2010ec}, which demonstrated that one can
have stability without supersymmetry. (It may be interesting to note
that the corresponding critical point in $\mathcal{N}=5$ supergravity
in four dimensions that breaks $SO(5)$ to $SO(3)$ actually was found
to be BF-stable already in the initial article by Breitenlohner and
Freedman on stability!)

Somewhat surprisingly, it turns out to be very difficult to obtain
this solution via the sensitivity backpropagation technique that
discovered most of the solutions found in the post-2000 era -- it
indeed had to be added to the list by hand. The reasons why the basin
of attraction of this particular solution in the numerical search is
so small is unclear at the time of this writing. Given that
sensitivity backpropagation using $|Q|^2$ as the objective function to
be minimized~\xcite{Fischbacher:2009cj} almost always seems to produce
either unstable non-supersymmetric or stable supersymmetric solutions
and at the time of this writing (version 1) {\em never} produced this
solution, this leaves a somewhat uneasy feeling about the possible
existence of (potentially a large number of) other stable
non-supersymmetric solutions that are not easily accessible by these
methods.

Still, while repeated numerical minimization seems to have a strong
chance of missing this solution, it indeed has been re-discovered
in~2002 using hybrid group-theoretic/numerical techniques on a certain
sub-manifold of ten $SO(3)$ invariant scalars described in the present
author's PhD thesis~\xcite{Fischbacher:2003rb}.

This critical point has the curious property that one can, at least in
the presentation given, triality-swap the $35_s$ entries with the
$35_c$ entries and again obtain the same solution. An attempt to
modify the numerical search in such a way that it looks for solutions
that also have this specific property was partially successful: the
modified search strategy did indeed manage to produce this solution
(in about 3\%~of all runs), as well as another solution that was
overlooked by the unmodified search, {\bf S0983994}. However, the
minima of the modified potential $|Q|^2+|\tilde Q|^2$ (where
$\tilde Q$ denotes the misalignment tensor for the triality-swapped
solution) typically are not also zeroes, and indeed {\bf S0983994}
does not have this triality-swapping property.

{\bf S1400056} is a novel unstable solution without any residual gauge
symmetry. It is noteworthy that, despite the very similar cosmological
constant, this is a different solution than {\bf S1400000},
with very different properties.

{\bf S1402217} is a novel unstable solution with residual gauge
symmetry $U(1)$.

{\bf S1424025} is a novel unstable solution with residual gauge
symmetry (likely) $SO(3)$.

{\bf S1441574} is a novel unstable solution with residual gauge
symmetry $U(1)$.

{\bf S1442018} is a novel unstable solution with residual gauge
symmetry $U(1)$.

{\bf S1443834} is a novel unstable solution with residual gauge
symmetry $U(1)$.

{\bf S1464498} is a novel unstable solution without any residual gauge
symmetry.

{\bf S1465354} is a novel unstable solution without any residual gauge
symmetry.

{\bf S1469693} is a novel unstable solution with residual gauge
symmetry $U(1)$.

\section{Conclusion}

Considering the effort that is normally required to obtain a new
critical point, the modified sensitivity backpropagation method, like
dynamite fishing, has a certain aura of being
unsportsmanlike. However, it is the result that counts, and quite
obviously, the data obtained through this powerful technique allows us
to develop a much better picture of spontaneous symmetry breaking in
$SO(8)$ supergravity than before.

It is amusing to note that obtaining new stationary points in the
scalar potential of $SO(8)$-gauged supergravity for a long time was
considered a hard problem. Now, it once again is.

\paragraph{Acknowledgments}

It is a pleasure to thank Nick Warner, Krzysztof Pilch, and Hermann
Nicolai for useful discussions. These were especially helpful for the
development of fast backpropagation-aware numerical code to calculate
the scalar mass matrix. Most calculations were done on the University
of Southampton's Iridis cluster.

\end{document}